\documentclass[10pt,aps,physrev,reprint,floatfix,superscriptaddress,longbibliography]{revtex4-2}
\pdfoutput=1
\usepackage{graphicx,latexsym}
\usepackage{dcolumn,amsfonts}
\usepackage{amssymb,amsmath,bm}

\usepackage[breaklinks=true,pdfencoding=auto]{hyperref}

\usepackage{natbib}
\hypersetup{
  colorlinks   = true, 
  urlcolor     = blue, 
  linkcolor    = blue, 
  citecolor    = red 
}

\usepackage{color}
\usepackage{ulem}

\begin{document}

\title{Unified approach to cyclotron and plasmon resonances in a periodic 2DEG\break
	   hosting the Hofstadter butterfly}

\author{Vidar Gudmundsson}
\email{vidar@hi.is}
\affiliation{Science Institute, University of Iceland, Dunhaga 3, IS-107 Reykjavik, Iceland}
\author{Vram Mughnetsyan}
\email{vram@ysu.am}
\affiliation{Department of Solid State Physics, Yerevan State University, Alex Manoogian 1, 0025 Yerevan, Armenia}
\author{Nzar Rauf Abdullah}
\affiliation{Physics Department, College of Science,
             University of Sulaimani, Kurdistan Region, Iraq}
\affiliation{Computer Engineering Department, College of Engineering, Komar University
             of Science and Technology, Sulaimani 46001, Kurdistan Region, Iraq}
\author{Chi-Shung Tang}
\email{cstang@nuu.edu.tw}
\affiliation{Department of Mechanical Engineering, National United University, Miaoli 36003, Taiwan}
\author{Valeriu Moldoveanu}
\email{valim@infim.ro}
\affiliation{National Institute of Materials Physics, PO Box MG-7, Bucharest-Magurele,
	        Romania}
\author{Andrei Manolescu}
\email{manoles@ru.is}
\affiliation{Department of Engineering, Reykjavik University, Menntavegur 1, IS-102 Reykjavik, Iceland}

%

\begin{abstract}
We present theoretical calculations for the cyclotron resonance and various
magnetoplasmon modes of a Coulomb interacting two-dimensional GaAs electron gas
(2DEG) modulated as a lateral superlattice of quantum dots subjected to an
external perpendicular constant magnetic field.
We use a real-time excitation approach based on the Liouville-von Neumann equation
for the density operator, that can go beyond linear response delivering
information of all longitudinal and transverse collective modes of interest to
the same order. We perform an extensive analysis of the coexisting
collective modes due to the lateral confinement and the magnetic field for
a different number of electrons in each dot. In the limit of vanishing dot modulation
of the 2DEG we find signs of the structure of the Hofstadter butterfly in
the excitation spectra.
\end{abstract}

\maketitle
%
%

\section{Introduction}
Besides their technological importance, two-dimen\-sion\-al electron systems in semi\-con\-ductor
het\-erostruc\-tures or quantum wells have served as important test beds for advancing
the understanding of quantum many-body methods and approaches in condensed matter theory
\cite{Ando82:437}.
Fundamental to this is the ability to change their electron density
or modulate it spatially into arrays of quasi-one or zero-dimensional electron systems.
The spatial and dynamical reduction of dimensions by external electric and magnetic
fields or microstructuring of the semiconductors modifies strongly the effective interactions
of the electrons in the systems and thus their electronic, optical and transport properties.

The Kohn’s theorem, published in 1961, states that the energy of the cyclotron resonance
of electrons in a uniform external magnetic field is independent of their mutual interactions
if the system is placed in a homogeneous rotating microwave field \cite{Kohn61:1242}.
In the early nineties of last century the this result was extended to explain
the two absorption lines appearing in FIR-infrared spectroscopy of parabolically confined quantum dots,
and the one line observed in quantum wires, in a constant external magnetic field \cite{Demel88:12732,Maksym90:108,Heitmann93:56,Shikin91:11903a}.
It was shown that in the dot system the two lines are due to the rigid oscillations of the center of
mass of the electron system with the rotation caused by the magnetic field, or against it.
The absorption frequencies are thus independent of the number of electrons. In the wire system there is
only a linear oscillation of the center of mass in the effective confining potential
renormalized by the magnetic field.

Subsequently, many research groups studied the effects of deviations from the parabolic
confinement or the circular shape in individual
dots \cite{Demel90:788a,Gudmundsson91:12098,Bollweg96:2774,Darnhofer96:591,PhysRevB.60.16591}
and wires \cite{Gudmundsson95:17744,Brataas96:4797}, or in
arrays of them \cite{Dempsey90:11708,Dahl92:15590,Kim93:11987,kotlyar98:3989,Zyl00:2107,Krahne01:195303},
just to mention few groups that modeled or measured the FIR-absorption of these
systems.

The 2DEG in a perpendicular constant magnetic field and a periodic lateral superlattice is known
for its fractal energy spectrum, the Hofstadter butterfly
\cite{Harper55:874,Azbel64:634,Langbein69:633,Hofstadter76:2239}. The screening of this spectrum
has been investigated at the Hartree level \cite{Gudmundsson96:5223R}, and its presence in the
FIR-infrared absorption of the system was investigated with this screening
included in the model \cite{Gudmundsson95:16744}.

The FIR-absorption of confined or periodically modulated electron systems has mostly been modeled
with the density-density response function of linear response \cite{Kubo57:570}, but
in order to capture the cyclotron resonance or more general transverse collective modes one needs to
resort to the current-current response function.
The cyclotron resonance has been explored for magneto polarons in parabolic quantum dots,
where Kohn's theorem is broken by the interactions with the phonon modes of
the lattice \cite{PhysRevB.47.12941},
or by interactions with fluctuations and impurities \cite{Merkt96:1134,CHEN20111007},
or in magnetically confined quantum dots \cite{PhysRevB.78.245311}.

Comparison of experimental results and models of the cyclotron resonance have shown the
importance of including many-body effects in the models \cite{PhysRevB.72.195328,PhysRevB.75.035334}.
Measurements of the cyclotron resonance in high mobility 2DEGs are known for
bringing into questions the properties of this fundamental excitation mode, especially
when interacting with other modes \cite{PhysRevB.67.241304}, or in more complex
experimental set-ups, with reflection spectroscopy in a terahertz band \cite{Kriisa2019},
or in combination with transport measurements in low density and mobility
samples under mm-wave irradiation \cite{PhysRevB.100.155301}.

A multitude of different approaches have been used to explore the time evolution
of electron systems subjected to short or periodic external excitations, but what
makes the linear response or the real-time excitation, described through the
L-vN equation, appealing is that in both approaches it is acknowledged that the
the external perturbation drives the electron system out of equilibrium.

Here, we want to present a unified approach, that can concurrently describe
the excitation of the longitudinal collective modes, the plasmons, and the transverse modes,
the cyclotron resonances and
transverse plasmons, in periodically laterally modulated 2DEG in a constant external
magnetic field. We will use the Liouville-von Neuman (L-vN) equation for a Hartree interacting
2DEG with a dot modulation, i.e.\ in a periodic square array of quantum dots, to investigate
the time-evolution of the system after it is excited with a short terahertz pulse with linear
or circular polarization.
The mathematical stability of this approach has been studied by Arnold et al.\
\cite{2003math.ph...5027A}, and it has been used to investigate various excitation spectra
of confined systems in magnetic field for excitation strength beyond linear response
\cite{Gudmundsson03:161301,ANDP:ANDP201400048}.

The external
magnetic field and the confinement potentials mix up the longitudinal and the transverse
collective modes and we will also explore their evolution as the modulation vanishes,
when the system changes from an array of quantum dots to a periodic 2DEG with vanishing
modulation.

We will compare the results for the density oscillations to corresponding results obtained
via the conventional linear response. Our ``real-time'' approach can furthermore be used to
access nonlinear response of the system, and as the excitation is with a short temporal pulse
it allows us to model terahertz pump-and-probe approaches.

In Section \ref{Model} we present the model for the static system, and show how the time
evolution of it is found after the excitation with an electrical pulse. In the Section
\ref{Results} we present the results of the calculations and make a comparison with
earlier approaches. In Section \ref{Conclusions} we draw our conclusions.

\section{Model}
\label{Model}

\subsection{Time-independent properties}
\label{Time-ind}
We explore the dynamics of electrons in a square superlattice of quantum dots
described by the potential
\begin{equation}
      V_\mathrm{per}(\bm{r}) = -V_0\left[\sin \left(\frac{g_1x}{2} \right)
        \sin\left(\frac{g_2y}{2}\right) \right]^2,
\label{Vper}
\end{equation}
with $V_0 = 16.0$ meV. The lattice is spanned by the lattice vectors
$\bm{R}=n\bm{l}_1+m\bm{l}_2$, with $n,m\in \bm{Z}$, and the unit vectors
are defined as $\bm{l}_1 = L\bm{e}_x$ and $\bm{l}_2 = L\bm{e}_y$.
The inverse lattice is spanned by $\bm{G} = G_1\bm{g}_1 + G_2\bm{g}_2$ with
$G_1, G_2\in \bm{Z}$, and the unit vectors are
\begin{equation}
      \bm{g}_1 = \frac{2\pi\bm{e}_x}{L}, \quad\mbox{and}\quad
      \bm{g}_2 = \frac{2\pi\bm{e}_y}{L},
\end{equation}
With $L = 100$ nm.

\begin{figure}[htb]
	\centerline{\includegraphics[width=0.48\textwidth]{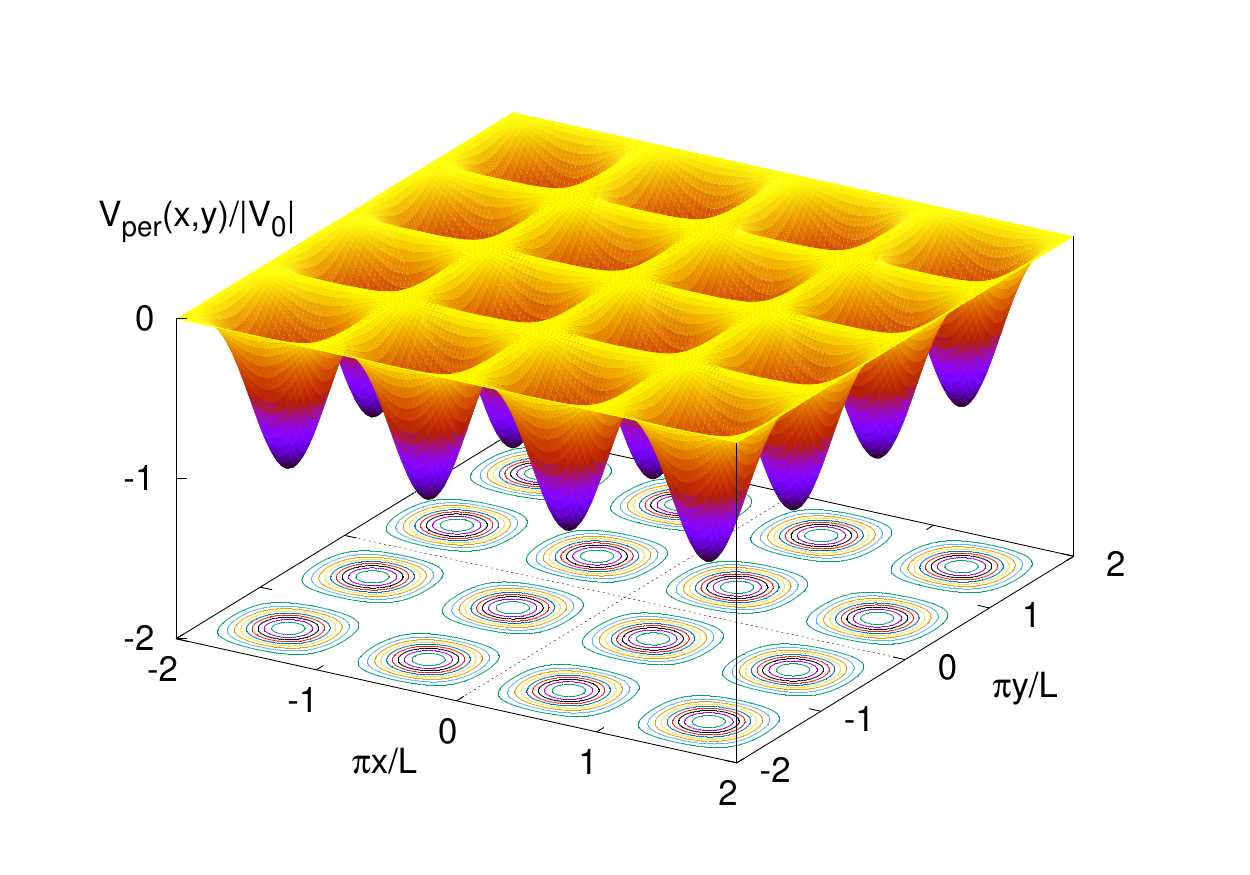}}
	\caption{The periodic potential $V_\mathrm{per}(\bm{r})$ (\ref{Vper}) describing
	         a square lattice of quantum dots with lattice length $L=100$ nm.}
	\label{Vxy}
\end{figure}

The Hamiltonian of the Hartree-interacting electrons in the periodic potential
(\ref{Vper}) and an external homogeneous perpendicular magnetic field is
\begin{equation}
      H = H_0 + V_\mathrm{H} + V_\mathrm{per},
\label{H}
\end{equation}
where $H_0$ is
\begin{equation}
      H_0 = \frac{1}{2m^*}\bm{\pi}^2, \quad\mbox{with}\quad
      \bm{\pi} = \left(\bm{p}+\frac{e}{c}\bm{A} \right).
\label{H0}
\end{equation}
We use a symmetric gauge for the vector potential,
$\bm{A}= (B/2)(-y,x)$
\cite{Ferrari90:4598,Silberbauer92:7355,Gudmundsson95:16744}
in order to make analytical calculations for the time dependent excitation
of the system more transparent analytically as the cartesian $x$- and $y$-coordinates
appear in a similar fashion in the natural eigenfunction basis introduced
below. The Hartree Coulomb interaction is
\begin{equation}
      V_\mathrm{H}(\bm{r}) = \frac{e^2}{\kappa}\int_{\bm{R}^2}d\bm{r}'\frac{\Delta n(\bm{r}')}
      {|\bm{r}-\bm{r}'|},
\label{Vcoul}
\end{equation}
with $\Delta n(\bm{r}) = n(\bm{r})-n_\mathrm{b}$, where $+en_\mathrm{b}$ is the
average background charge density needed to keep the total system charge neutral
(the average charge density of the ions in the crystal lattice). For GaAs we assume
$\kappa = 12.4$, and the effective mass $m^* = 0.067m_e$. The mean number of electrons
in each dot is noted by $N_e$.

The external homogeneous magnetic field $\bm{B}=B\bm{e}_z$ imposes a length scale
on the 2DEG in the $x$-$y$-plane, the magnetic length $l=\{\hbar c/(eB)\}^{1/2}$.
At the same time the 2DEG is also under the influence of the lattice length
$L$ of the square superlattice. The translation operator
$T({\bm{R}})=\exp(-i\bm{R}\cdot\bm{p}/\hbar )$ does not commute with $H_0$,
$[T(\bm{R}),H_0]\neq 0$, but the magnetic translation operator
$S(\bm{R})=\exp \{ie/(\hbar c)\chi\}T(\bm{R})$ with $\chi$ obtained from
$T^{-1}\bm{A}T = \bm{A}+\bm{\nabla}\chi$ fulfills
\begin{align}
      S^{-1}(\bm{R})\bm{\pi}S(\bm{R}) &= \bm{\pi}\nonumber\\
      S^{-1}(\bm{R})\bm{r}S(\bm{R}) &= \bm{r} + \bm{R}
\end{align}
and thus $[S(\bm{R}),H_0]=0$ and $[S(\bm{R}),H]=0$. Nevertheless,
generally $[S(\bm{R}_1),S(\bm{R}_2)]\neq 0$, unless, an integer number
of flux quanta $\Phi_0 = hc/e$ flows through the lattice unit cell.
This reflects the incommensurability of the two length scales, $l$ and $L$.
We follow Ferrari \cite{Ferrari90:4598} and Silberbauer \cite{Silberbauer92:7355}
introducing a sublattice $\bm{c}=\bm{l}_1/p$ and $\bm{d}=\bm{l}_2/q$ with
$p,q\in\bm{N}$, where only one flux quantum $\Phi_0$ flows through the primitive
unit cell of the sublattice, to construct the eigenfunctions of $H_0$
\begin{equation}
      \phi_{n_l}^{\mu\nu}(\bm{r}) = \frac{1}{\sqrt{pq}}\sum_{n,m=-\infty}^\infty
      [S(\bm{c})e^{-i\mu}]^m [S(\bm{d})e^{-i\nu}]^n \phi_{n_l}(\bm{r}),
\label{phi_munu}
\end{equation}
where
\begin{equation}
      \phi_{n_l}(\bm{r}) = \frac{1}{\sqrt{2\pi l^2 n_l!}}
      \left(\frac{x+iy}{\sqrt{2}l} \right)^{n_l}
      \exp\left(-\frac{r^2}{4l^2} \right),
\label{phi_nl}
\end{equation}
with the Landau level number $n_l = 0,1,2,\dots$ and
$\mu = (\theta_1+2\pi n_1)/p$, $\nu = (\theta_2+2\pi n_2)/q$,
where $n_1\in I_1 = \{0,\dots, p-1 \}$, $n_2\in I_2 = \{0,\dots, q-1 \}$,
and $\theta_i\in [-\pi,\pi]$. The eigenfunctions of $H_0$ (\ref{phi_munu})
form a complete orthonormal Hilbert space ${\cal H}_{\theta_1\theta_2}$,
if $(\mu,\nu )\neq (\pi,\pi )$ for all $(n_1,n_2)\in I_1\times I_2$.
The eigenfunctions (\ref{phi_munu}) have to be normalized on a
primitive unit cell of the direct lattice with
\begin{align}
      ||\phi_{n_l}^{\mu\nu}(\bm{r})||^2 = \sum_{n,m=-\infty}^\infty
      (&-1)^{mn}e^{i(\mu m+\nu n)}\nonumber\\
      &\times\exp \left(-\frac{1}{4l^2}|m\bm{d}+n\bm{c}|^2\right).
\end{align}

The states of the Hamiltonian $H$ (\ref{H}) are determined self-consistently
within the Hilbert space of the orthonormal eigenfunctions of $H_0$ (\ref{H0})
with the condition that they remain orthogonal throughout the iterations
in accordance with the usual methodology applied in the Hartree-approximation.

The primitive unit cell in the direct lattice will be noted by ${\cal A}$.
Its area will also be noted by ${\cal A}=L^2$ for the square lattice and
the number of magnetic flux units through it is $pq = B{\cal A}/\Phi_0$.

\subsection{Real-time excitation}
\label{Time-dep}
For the time-independent mean-field description of electrons in a two-dimensional
doubly periodic problem we used a basis of states
$|\bm{\alpha}\rangle =|\mu,\nu,n_l\rangle$
living in each point $(\theta_1,\theta_2)$ of the magnetic Brillouin zone (MBZ).
Due to the properties of the electron-electron Coulomb interaction the same could
be stated for the interacting mean-field states $|\bm{\alpha})$.
Note here that $\langle\bm{r}|\bm{\alpha}\rangle = \phi_{n_l}^{\mu\nu}(\bm{r})$.
An external time-dependent electric field pulse with circular polarization,
or its potential
\begin{align}
      V^\mathrm{ext}(\bm{r,t}) = V_\mathrm{t} \left\{ (\Gamma t)^2 e^{-\Gamma t}\right\}
      [ &\cos{(k_y y)}\cos{(\Omega t)} \nonumber \\
                + c_r&\cos{(k_x x)\sin{(\Omega t)}} ]
\label{phi-ext}
\end{align}
with $c_r=\pm 1$
breaks this symmetry and mixes up states at different points,
$(\theta_1,\theta_2)$, in the MBZ, as the wavevectors $k_x$ and $k_y$
are generally neither commensurate with $L$ nor $l$.
We need a larger Hilbert space with states $|\bm{\alpha\theta} )$ with
$\bm{\theta} = (\theta_1,\theta_2)$, and as the periodic 2DEG is infinite
in extent, we have a Hilbert space of continuous states grouped into
discrete energy bands, a rigged Hilbert space \cite{Madrid_2005}.

As the eigenfunctions of $H_0$ (\ref{H0}), $\phi_{n_l}^{\mu\nu}$, and $H$ (\ref{H}),
$\psi_{\bm{\alpha}}$, both live
at a definite point in the MBZ we can express their translation
as
\begin{align}
      \psi^*_{\bm{\alpha\theta}}&(\bm{x+R})\psi_{\bm{\beta\theta}'}(\bm{x+R})
      \nonumber\\ =& \{e^{in(\theta_1-\theta'_1)+im(\theta_2-\theta'_2)} \}
      \psi^*_{\bm{\alpha\theta}}(\bm{x})\psi_{\bm{\beta\theta}'}(\bm{x}),
\label{psipsi}
\end{align}
where we add a reference to the $\bm{\theta}=(\theta_1,\theta_2)$ point of the
MBZ. The Hartree-interaction (\ref{Vcoul}) does not break this symmetry as it is only a
functional of the periodic electron density.
We will now consider the wavefunction $\psi_{\bm{\alpha\theta}}(\bm{r}) =
\langle\bm{r}|\bm{\alpha\theta})$ and define the inner product of two Hartree-interacting
states in the periodic potential and the extended Hilbert space as
\begin{align}
      (\bm{\alpha\theta}&|\bm{\beta\theta}') =
      \int_{\bm{R}^2} d\bm{r}\: \psi^*_{\bm{\alpha\theta}}(\bm{r})
      \psi_{\bm{\beta\theta}'}(\bm{r}) \nonumber\\
      &= \frac{(2\pi)^2}{\cal A}\delta^G(\bm{\theta}-\bm{\theta}')
      \int_{\cal A} d\bm{r}\: \psi^*_{\bm{\alpha\theta}}(\bm{x})\psi_{\bm{\beta\theta}'}(\bm{x}) ,
\label{innerproduct}
\end{align}
where the integral over the entire $\bm{R}^2$ has been folded back into
the primitive unit cell of the lattice, and $\delta^G(\bm{\theta}-\bm{\theta}')$
is the Dirac $\delta$-function periodic with respect to the inverse lattice.
Here, we use the notation that $\bm{r}\in\bm{R}^2$, but $\bm{x}\in\cal{A}$.

The coupling of states between different points in the MBZ also
emerges when linear response formalism is used to calculate the
response of the system to an external excitation with finite wavevector
\cite{Gudmundsson96:5223R,Dahl93:15480,Dahl90:5763}, but the straight
forward structure of the response function needed can be expressed
without a construction of a larger Hilbert space. Here, we want to
be able to go beyond a linear response formalism \cite{Kubo57:570} by using, without
an approximation, the Liouville-von Neumann equation (L-vNE)
\begin{equation}
      i\hbar\partial_t\rho(t) = \left[H[\rho (t)],\rho (t)\right].
\label{L-vNE}
\end{equation}
It is thus convenient to express the density operator $\rho$ as a matrix in a
larger Hilbert space.
The (L-vNE) (\ref{L-vNE}) has been used to investigate strong excitation of individual quantum
dots in comparison to other methods \cite{Gudmundsson03:161301}.

The local electron density is evaluated via the density operator through
\begin{align}
      n(\bm{r},t)& = \mathrm{Tr}\{\delta (\hat{\bm{r}}-\bm{r})\rho (t) \} =\nonumber\\
      &\frac{1}{(2\pi)^4}\int_{-\pi}^\pi d\bm{\theta}d\bm{\theta}'
      \sum_{\alpha\beta}\psi^*_{\bm{\alpha\theta}}(\bm{r})
      \psi_{\bm{\beta\theta}'}(\bm{r})\rho_{\bm{\beta\theta}',\bm{\alpha\theta}}(t),
\label{Ne}
\end{align}
where the off-diagonal elements in the density operator play an essential role
as their contribution to the density conveys the symmetry breaking effects of the
external potential (\ref{phi-ext}) on the 2DEG. The length scale imposed by the
external excitation potential (\ref{phi-ext}) breaks the symmetry imposed by
the two commensurate scales, the magnetic
length, $l$, and the lattice length $L$ of the square lattice.

The Hamiltonian in the L-vNE (\ref{L-vNE}) is the time independent Hamiltonian
(\ref{H}) with a time dependent term added
\begin{equation}
      H(t) = H_0 + H_\mathrm{H} + V_\mathrm{per} + H_\mathrm{I}(t),
\label{Ht}
\end{equation}
where $H_\mathrm{I}(t)$ stands both for the time-dependent external potential (\ref{phi-ext})
and the residual Coulomb potential, the addition to the Hartree potential stemming from the
self-consistent changes to the electron density inflicted by the time dependent external
excitation. This dynamical correction to the Hartree potential is now time-dependent due to it
being a functional of the time-dependent density operator $\rho (t)$, see Appendix
\ref{AppNumDetails}. The L-vNE is then
\begin{align}
      i\hbar&\partial_t\rho_{\bm{\alpha\theta},\bm{\beta\theta}'}(t) =
      \left\{E_{\bm{\alpha\theta}} - E_{\bm{\beta\theta}'} \right\}
      \rho_{\bm{\alpha\theta},\bm{\beta\theta}'}(t)\nonumber\\
      &+ \frac{1}{(2\pi)^2}
      \sum_\gamma \int_{-\pi}^\pi d{\bm{\theta}}''
      \left\{(\bm{\alpha\theta}\left|H_I[\rho]\right|\bm{\gamma\theta}'')
      \rho_{\bm{\gamma\theta}'',\bm{\beta\theta}}(t)\right.\nonumber\\
      &\qquad\qquad\qquad\qquad -
      \left.\rho_{\bm{\alpha\theta},\bm{\gamma\theta}''}(t)
      (\bm{\gamma\theta}''\left|H_I[\rho]\right|\bm{\beta\theta}') \right\}.
\label{L-vNE2}
\end{align}
To evaluate the matrix elements in the last terms of the L-vNE (\ref{L-vNE}) we
need the translation properties of the wavefunctions (\ref{psipsi}) and the
residual Hartree interaction
\begin{equation}
      \delta V_H(\bm{r},t) = \frac{e^2}{\kappa}\int_{\bm{R}^2}d\bm{r}'\:\frac{\delta n(\bm{r}',t)}
      {|\bm{r}-\bm{r}'|},
\label{Vcoult}
\end{equation}
with
\begin{align}
      \delta n(\bm{r},t) = &\: \Delta n(\bm{r},t) - \Delta n(\bm{r},0)\nonumber\\
                         = &\:	      n(\bm{r},t) -        n(\bm{r},0),
\label{dn}
\end{align}
but here we can not assume $\delta n(\bm{r},t)$ to have the
same periodicity as the lattice, like was possible for the time-independent static system.

The translation properties of the wavefunctions allows the integral over $\bm{R}^2$
to be folded back to an integral over the unit cell ${\cal A}$ and the matrix elements
for the Hartree term become
\begin{align}
      (\bm{\mu\theta}|V_H(t)&|\bm{\nu\theta}') = \frac{1}{\cal A}\sum_{\bm{G}}
      V_H(\bm{G}+\tilde{\bm{\theta}}-\tilde{\bm{\theta}}',t)\nonumber\\
      \int_{{\cal A}}& d\bm{x}\;
      e^{i(\bm{G}+\tilde{\bm{\theta}}-\tilde{\bm{\theta}}')\cdot\bm{x}}
      \psi^*_{\bm{\mu\theta}}(\bm{x})\psi_{\bm{\nu\theta}'}(\bm{x}),
\label{V_H}
\end{align}
where $\tilde{\bm{\theta}} = (\theta_1/l_1,\theta_2/l_2)$,
$\tilde{\bm{k}} = (k_1l_1,k_2l_2)$, and
\begin{align}
      V_H(\bm{G}+\bm{k},t)& =
      \frac{2\pi e^2}{\kappa |\bm{G}+\bm{k}|}\nonumber\\
      \frac{1}{(2\pi )^2{\cal A}} &\sum_{\alpha\beta}\int_{-\pi}^{\pi}d\bm{\theta}
      \int_{{\cal A}} d\bm{x}\; e^{-i(\bm{G}+\bm{k})\cdot\bm{x}}\nonumber\\
      &\qquad  \psi^*_{\bm{\alpha\theta}}(\bm{x})
      \psi_{\bm{\beta\theta}+\tilde{\bm{k}}}(\bm{x})
      \Delta\rho_{\bm{\beta\theta}+\tilde{\bm{k}},\bm{\alpha\theta}}(t).
\label{VHGk}
\end{align}
In Eq.\ (\ref{VHGk}) a general wavevector $\bm{q} = \bm{G} + \bm{k}$ has been written in terms
of a vector in the reciprocal lattice, $\bm{G}$, and a residual $\bm{k}$ wavevector
residing inside the first MBZ (see Appendix \ref{AppNumDetails} about the numerical
implementation).

The matrix elements of the external potential (\ref{phi-ext}) are calculated with an integral
over the entire $\bm{R}^2$ with the same backfolding into the unit cell as was employed in
Eq.\ (\ref{V_H}) for the matrix elements of the Coulomb potential (\ref{Vcoult}).
To analyze the effects of the excitation on the periodic 2DEG we calculate the
time-dependent induced electron density $\delta n(\bm{r},t) = n(\bm{r},t) -
n(\bm{r},0)$ and the averages $\langle\hat{O} \rangle = \mathrm{Tr}\{\hat{O}\rho (t)\}$
for the dipole operators $\hat{O} = \hat{x}$ and $\hat{O} = \hat{y}$, the quadrupole operator
$\hat{O} = \hat{yx}-\langle\hat{y}\rangle\langle\hat{x}\rangle$, and the monopole operator
$\hat{O} = \hat{x}^2+\hat{y}^2-\langle\hat{x}\rangle^2-\langle\hat{y}\rangle^2$, where
the matrix elements of $\hat{O}$ are evaluated with an spatial integral over just one
unit cell. Defined in this way the last two operators are the relative quadrupole
and monopole operators for the case of $N_e=1$ in an individual isolated
quantum dot \cite{Puente01:235324}.
The average for the quadrupole operator we indicate with $Q_2$ and use $Q_0$ for
the monopole operator. All the aforementioned averages
gauge collective modes with density variations, in order to detect rotational collective
modes (transverse modes) we need to gauge modes, where the current density
$\bm{j}=-e\dot{\bm{r}} = -(ie/\hbar)[H(t),\bm{r}]$ plays a role \cite{Gudmundsson03:161301}.
This will be accomplished by calculating the dimensionless quantity
\begin{equation}
      Q_{\bm{j}} = \frac{1}{l^2\omega_c}
      \langle i(\bm{r}\times\dot{\bm{r}})\cdot\hat{\bm{z}}\rangle ,
\label{Qj}
\end{equation}
which is directly proportional to the orbital part of the magnetization measured
in one cell, or the orbital angular momentum in the cell.

\section{Results}
\label{Results}
For the ``real-time'' excitation with a linear polarization we use the external
potential
\begin{equation}
      V^\mathrm{ext}(\bm{r,t}) = V_\mathrm{t} e^{-\Gamma t}
      \cos{(k_i x_i)}\sin{(\Omega t)},
\label{phi-ext-lin}
\end{equation}
where $k_ix_i = k_xx$ or $k_yy$ depending on whether the polarization is along
the $x$- or the $y$-direction in the plane of the 2DEG.
Originally (\ref{phi-ext-lin}) was chosen for the excitation with linear polarization,
and modified and extended to circular polarization as (\ref{phi-ext}) in order to
have the excitation always starting from 0 at $t=0$.

To understand the nature of the confining dot potential we can consider a
polar coordinate system with origin at the minimum of the periodic potential
(\ref{Vper}) in one cell. An expansion to the fourth order around the minimum gives
\begin{align}
      V_\mathrm{per}(x,y)\approx& V_0 \left[
      \left(\frac{\pi^2}{L^2}\right) r^2 \right.  \nonumber\\
      &\left. -\left(\frac{\pi^4}{12L^4}\right) r^4\left\{1+\frac{3\sin^2(2\theta)}{2}\right\}
      -1\right] ,
\label{VperApp}
\end{align}
if $r << L$.
The parabolic part of the expansion (\ref{VperApp}) would lead to a confinement
energy $\hbar\omega_0\approx 6.0$ meV. As expected the forth order term describes
a weakening of the confinement potential, and a square symmetric deviation from the
circular shape at low energy. As the energy is increased the shape of the
square lattice takes over.
\begin{figure}[htb]
	\centerline{\includegraphics[width=0.48\textwidth]{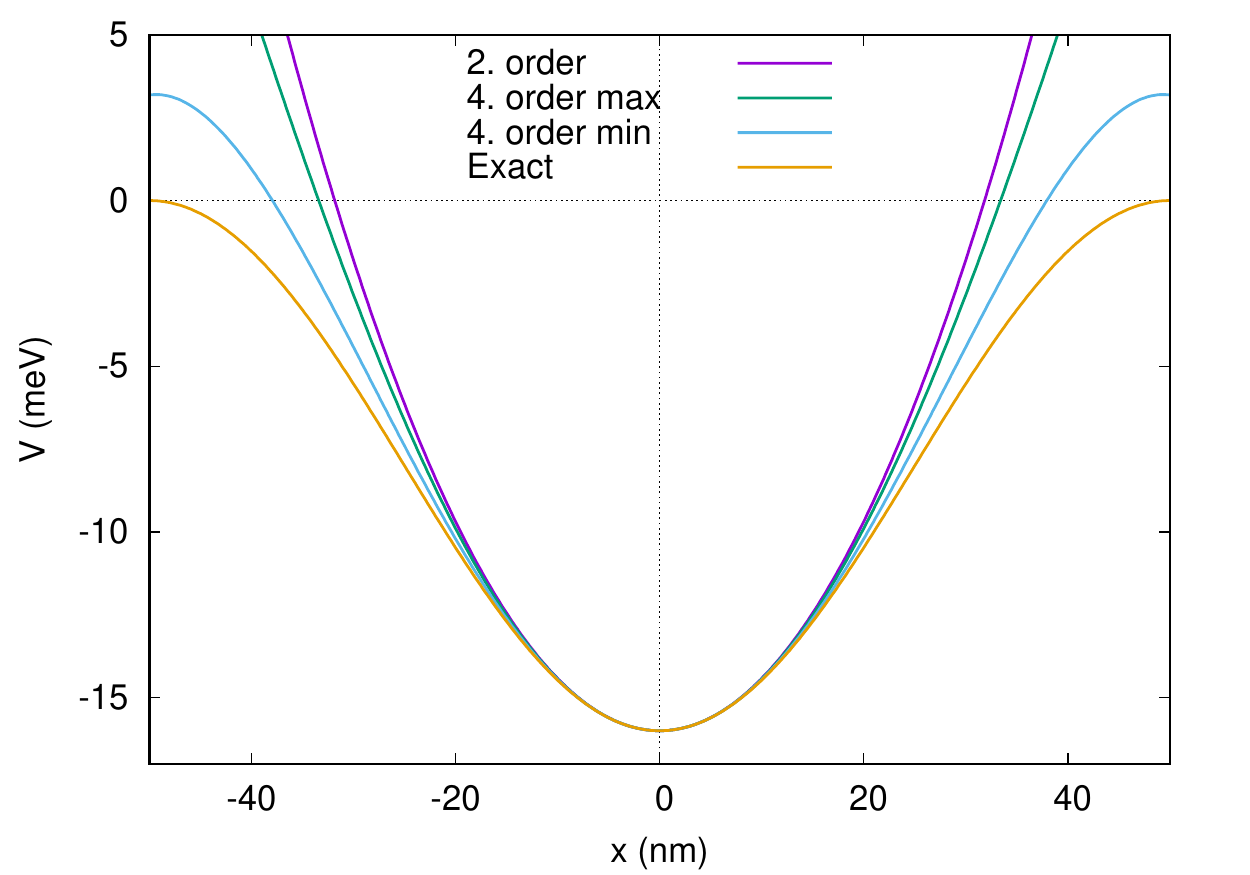}}
	\caption{The confinement potential defining the dot in each unit cell
	         as seen along the $x$-axis close to their center. Compared are the exact potential
	         (orange), the second order expansion from Eq.\ (\ref{VperApp}) (violet),
	         and the fourth order (green) along  $\theta = 0+n\pi$, and the forth order along
	         $\theta = \pi/4+n\pi$ (blue), with $n=0,1,2,3\dots$.
	         The difference between the max ($\theta = 0+n\pi$) and the min ($\theta = \pi/4+n\pi$)
	         fourth order expansions gives an indication of the deviation from
	         circular symmetry imposed by the square lattice.}
	\label{VxyApprox}
\end{figure}

If the magnetic field is increased (the number of flux quanta through the unit cell
$pq$) the magnetic length $l$ becomes smaller compared to the lattice length $L$
and an electron in the lowest state becomes better localized in the dot potential
(\ref{Vxy}). The magnetic length $l$ is then a convenient parameter to compare to
the potential in Fig.\ \ref{VxyApprox} in order to determine the degree of deviation
from parabolic confinement felt by an electron.

\subsection{Comparison to linear response}
To compare to the results of the real-time excitation of the system we use the linear
response model developed earlier for the absorption $P(\omega )$ \cite{Gudmundsson96:5223R}
using $n_l=10$ as the number of Landau Levels and a $32\times 32$ grid of unevenly spaced $(\theta_1,\theta_2)$-points for the unit cell
in the reciprocal lattice for a repeated 4-point Gaussian quadrature. In order to smear
out the effects of the singularities in the response functions we use a constant broadening
$\hbar\eta=0.2\hbar\omega_c$. For the real-time excitation we employ an 8x8 equally spaced grid
in the reciprocal unit cell in combination with a repeated Booles quadrature and no level
broadening. Furthermore, we use $n_l=10$ for the static part of the calculation, but limit the
number of Landau bands used in the
calculation of the time-evolution to $n_\mathrm{Ht}=8$. Here we are not aiming at exploring the
response of the system far beyond the linear response regime, and we have tested the convergence
of the results for the strength selected for the excitation.

\begin{figure}[htb]
	\centerline{\includegraphics[width=0.47\textwidth,bb = -30 60 395 300]{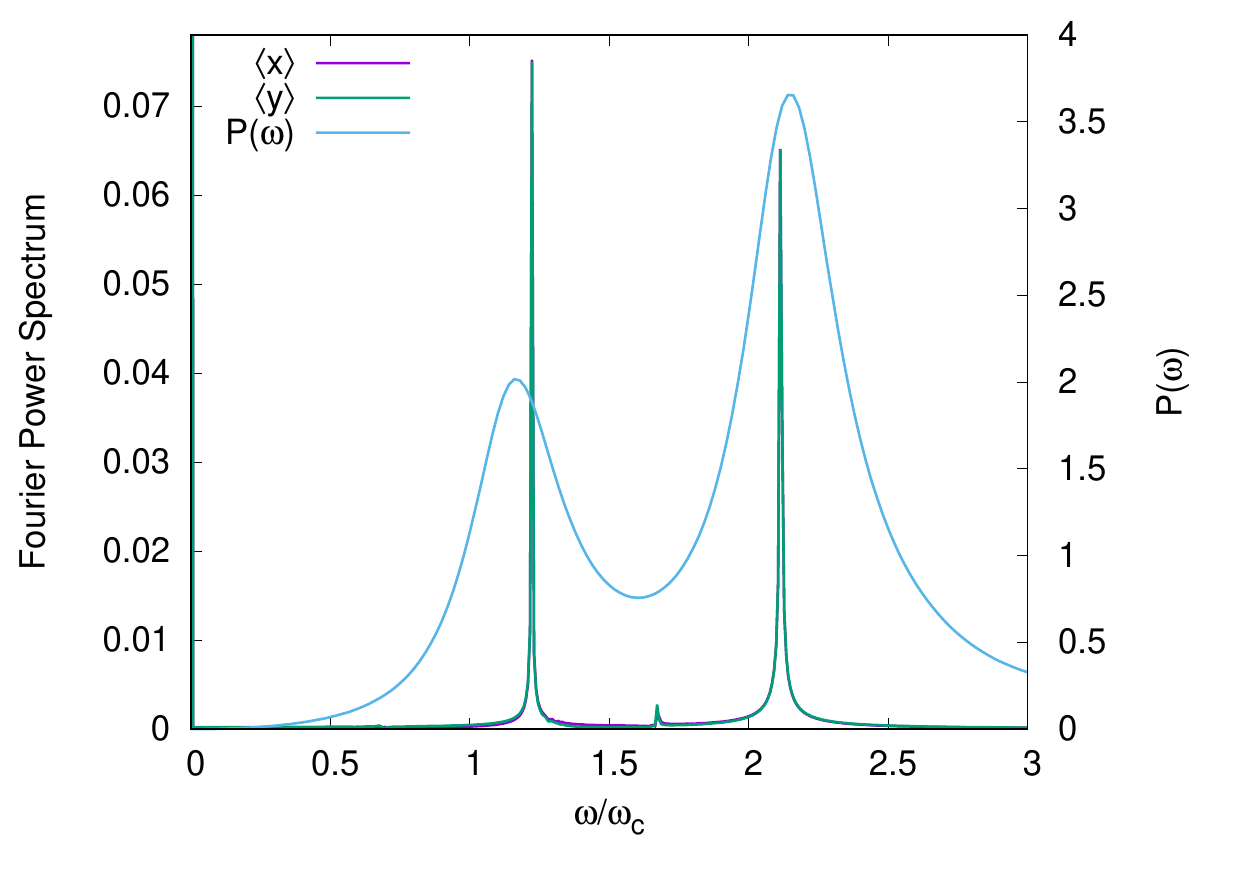}}
	\centerline{\includegraphics[width=0.47\textwidth,bb = -30 60 395 290]{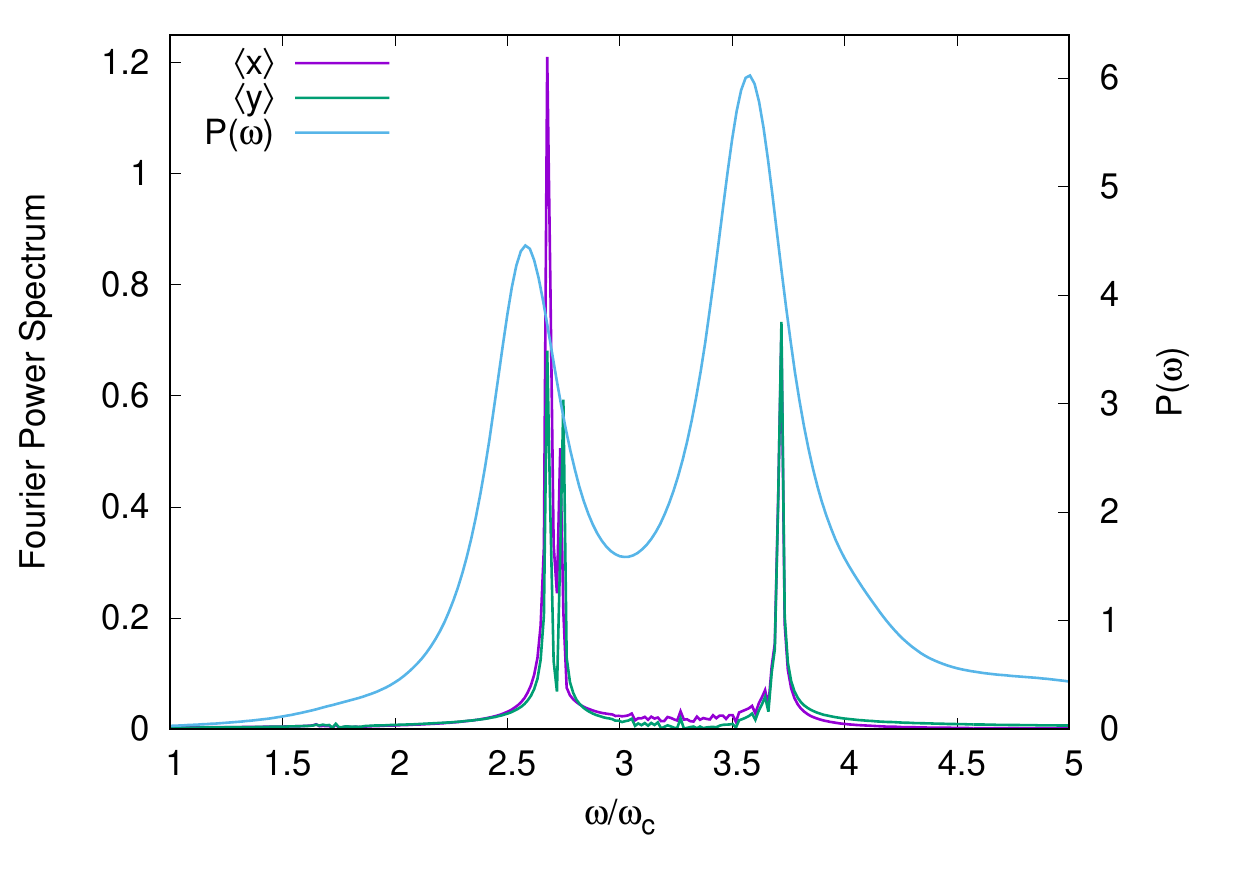}}
	\centerline{\includegraphics[width=0.47\textwidth,bb = -30 20 395 290]{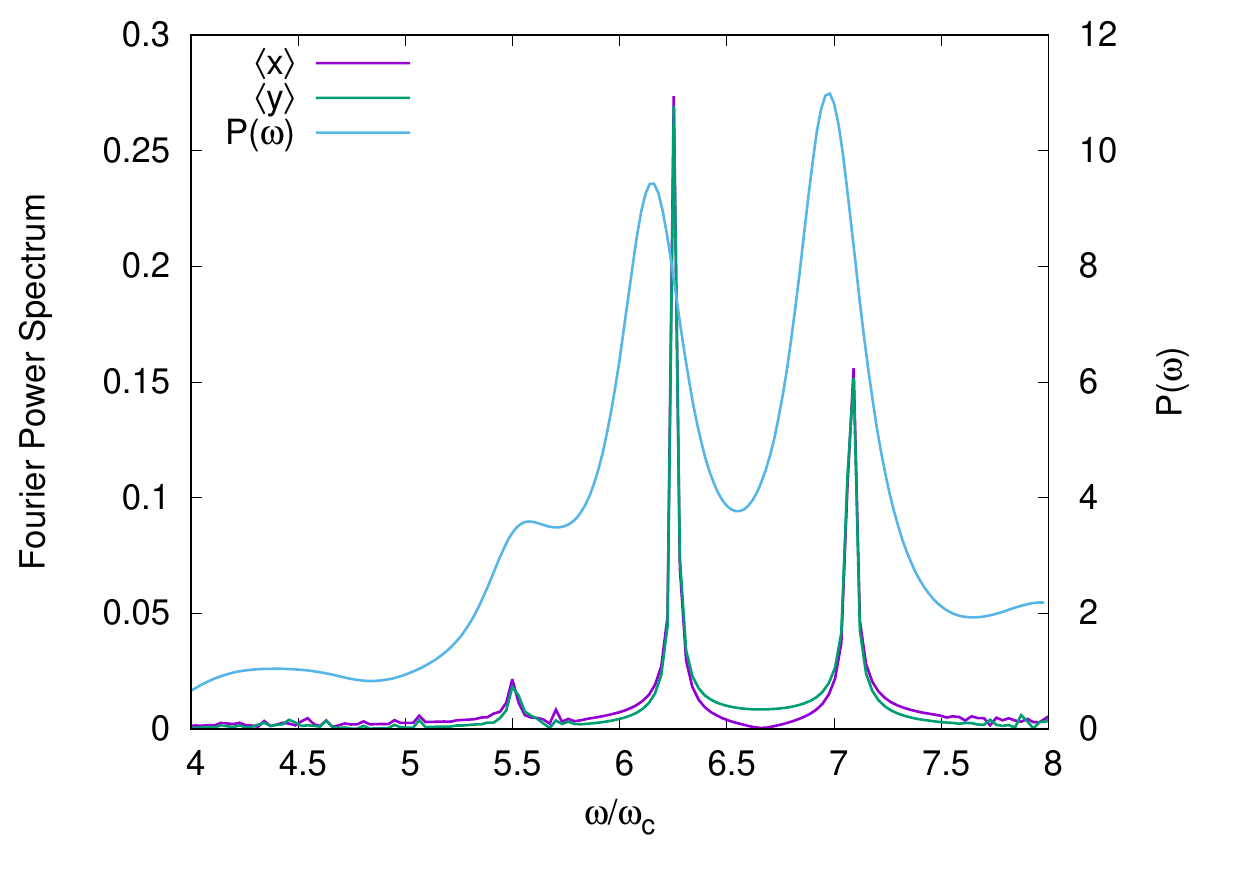}}
    \caption{Comparison of the Fourier power spectra for $\langle x\rangle$ and $\langle y\rangle$
             for linearly polarized excitation of a rectangular array of quantum dots with the
             linear response absorption $P(\omega )$ for the same system for $pq=4$ (top),
             $pq=2$ (center), and $pq=1$ (bottom). In the linear response calculation the broadening in the
             response function is $\hbar\eta = 0.2\hbar\omega_c$ and $k_xL=k_yL=0.01$.
             In the real-time excitation $k_xL=0.0001$ $k_yL=0$ and $V_\mathrm{t}=0.01$ meV
             for $pq=4$ and $pq=1$, but $V_\mathrm{t}=0.1$ meV for $pq=2$.
             $N_e=1$, $V_0=-16$ meV, $L=100$ nm, $T=1.0$ K, $\hbar\Gamma = 4.0$ meV, and
             $\hbar\Omega = 4.5$ meV.}
    \label{Compare}
\end{figure}

In Fig.\ \ref{Compare} we compare $P(\omega)$ for dipole excitations in the linear response model
with the Fourier power spectrum for the mean values $\langle x\rangle$ and $\langle y\rangle$
for the real-time excitation for three values of the magnetic flux $pq$ through the unit cell.
In all the calculations we use linear polarization of the excitation with a very low impulse $k_xL$ or
$k_yL$ in order to explore the Kohn modes for one electron in each quantum dot in the array
\cite{Kohn61:1242,PhysRevB.42.1486,Maksym90:108,Gudmundsson91:12098}. No significance should be given to the
change in relative height of the main peaks in the real-time excitation as that simply depends
on the frequency distribution offered to the system by the shape of the excitation pulse
used. Here, we have taken care of not including too high frequency in the pulse.
In the real-time excitation no broadening is assumed, but the length of the time series
gives the peaks in the Fourier power spectrum an apparent width. With few exceptions we
have calculated the time series with 10200 steps of length 0.02 ps each.

The magnetic length $l\approx 19.9$ nm for $pq=4$, 28.2 nm for $pq=2$, and 39.9 nm for $pq=1$,
and consequently, though not shown here, the overlap of the electron density is very small for
$N_e=1$ at $pq=4$, but considerable for $pq=1$, and not large, but not negligible for $pq=2$.
Accordingly, we see in Fig.\ \ref{Compare} for $pq=4$ (upper panel) rather clean two Kohn peaks.
For $pq=2$ (center panel) we see two Kohn peaks, but the lower one shows a small splitting
caused by a slight influence of the square symmetry of the lattice at this magnetic flux
\cite{PhysRevB.60.16591}. At still lower flux $pq=1$ (bottom panel) the splitting of the
lower peak is clearly visible.

Not shown here, but a graph of $\langle y\rangle$ versus $\langle x\rangle$ shows an extremely
simple pattern for the (center of mass) CM-motion known for circular parabolically
confined quantum dots
for $pq=4$, but with some slight modulation for $pq=3$, that is increased for $pq=2$.
For $pq=1$ the simplicity is lost, but the pattern is still very regular.

For collective density modes comprised of dipole active transitions in arrays of quantum
dots in a magnetic field the real-time excitation method results in higher resolution of
the mode spectrum as no broadening has to be assumed for the underlying single-electron
states of the system. The total energy is strictly conserved in the ``real-time'' excitation
after the pulse has died out, but in linear response nothing can be stated about the energy
conservation since the results are linear in the excitation potential.

A slightly different approach ``to real-time'' excitation of a single
quantum dot with Hartree-Fock interacting electrons has been
compared to results obtained via linear response for the corresponding system
\cite{Valin-Rodriguez2002,Puente01:235324}.

\subsection{Extended mode spectrum for quantum dots}
Deviations from circular shape and parabolic confinement of individual dots in the array
and excitation of the system with pulses that carry a wavevector supplying an impulse
to it makes it important to investigate the mode spectrum in the relevant energy range.
The linear response built on the density-density response function can supply information
about dipole and quadrupole collective density modes and in some special situations the monopole
or breathing mode (longitudinal modes), but the external magnetic
field can lead to collective current modes (transverse modes) that need to be examined through
the current-current response function. In the real-time excitation method all these modes can become
active and their ``detection'' depends on the expectation values of which operators are registered
through the time evolution of the system.

\begin{figure*}[htb]
	{\includegraphics[width=0.47\textwidth]{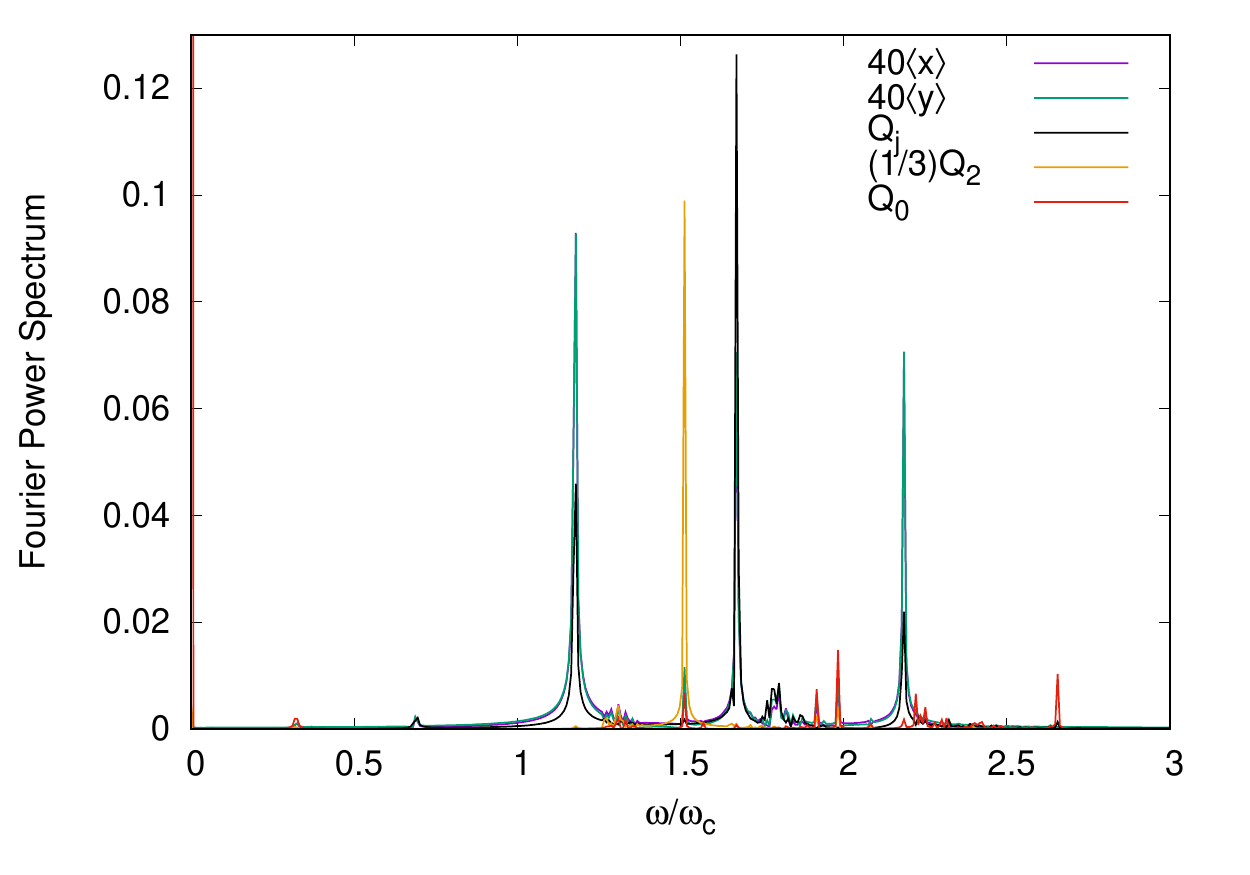}}
	{\includegraphics[width=0.47\textwidth]{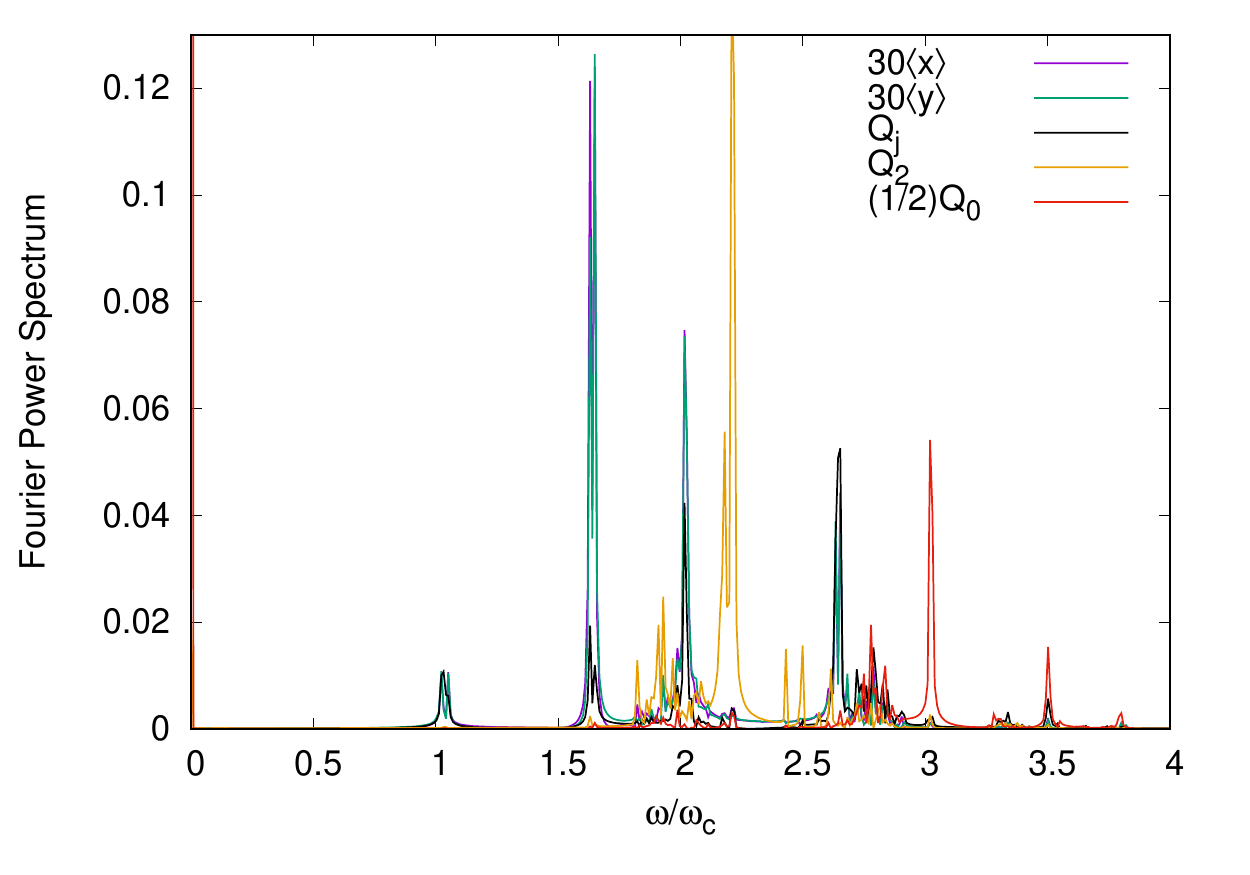}}\\
	{\includegraphics[width=0.47\textwidth]{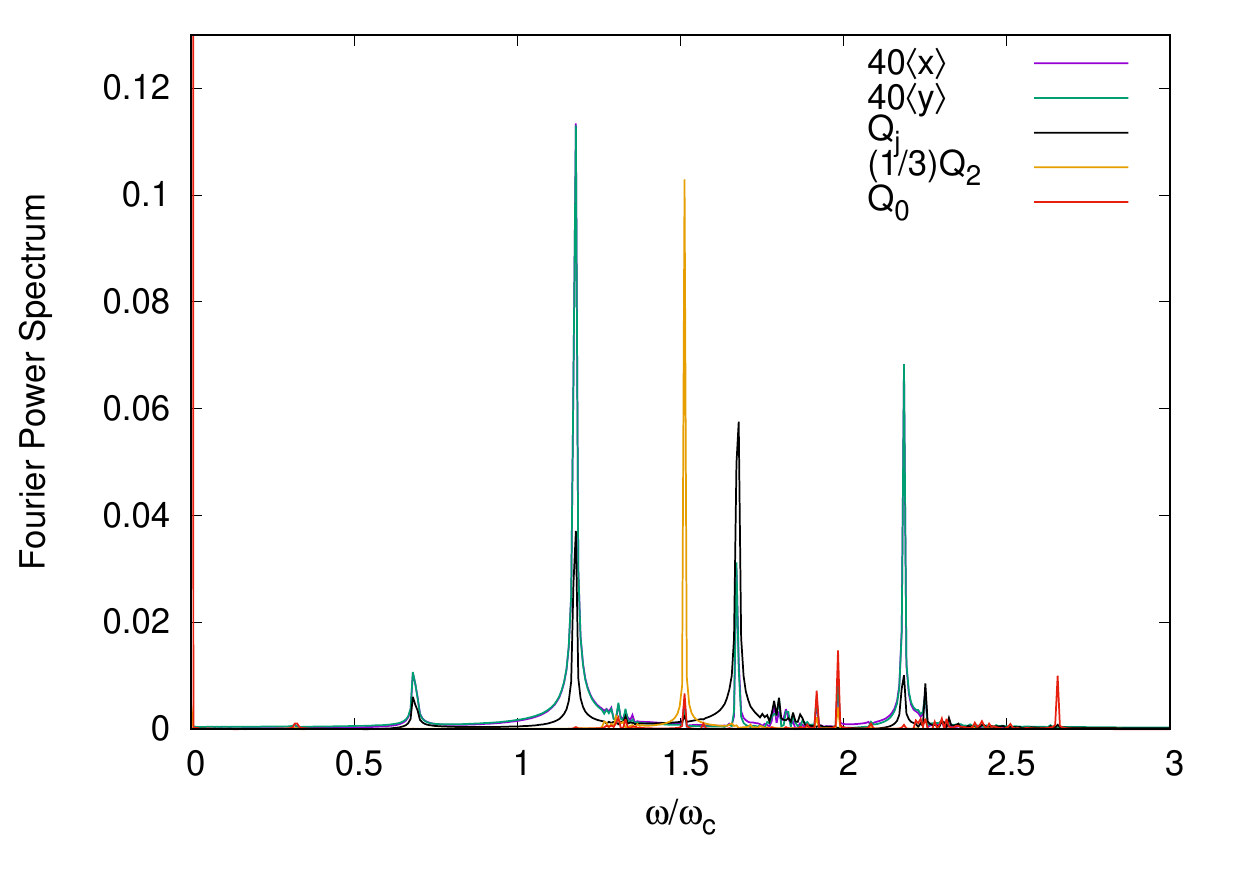}}
	{\includegraphics[width=0.47\textwidth]{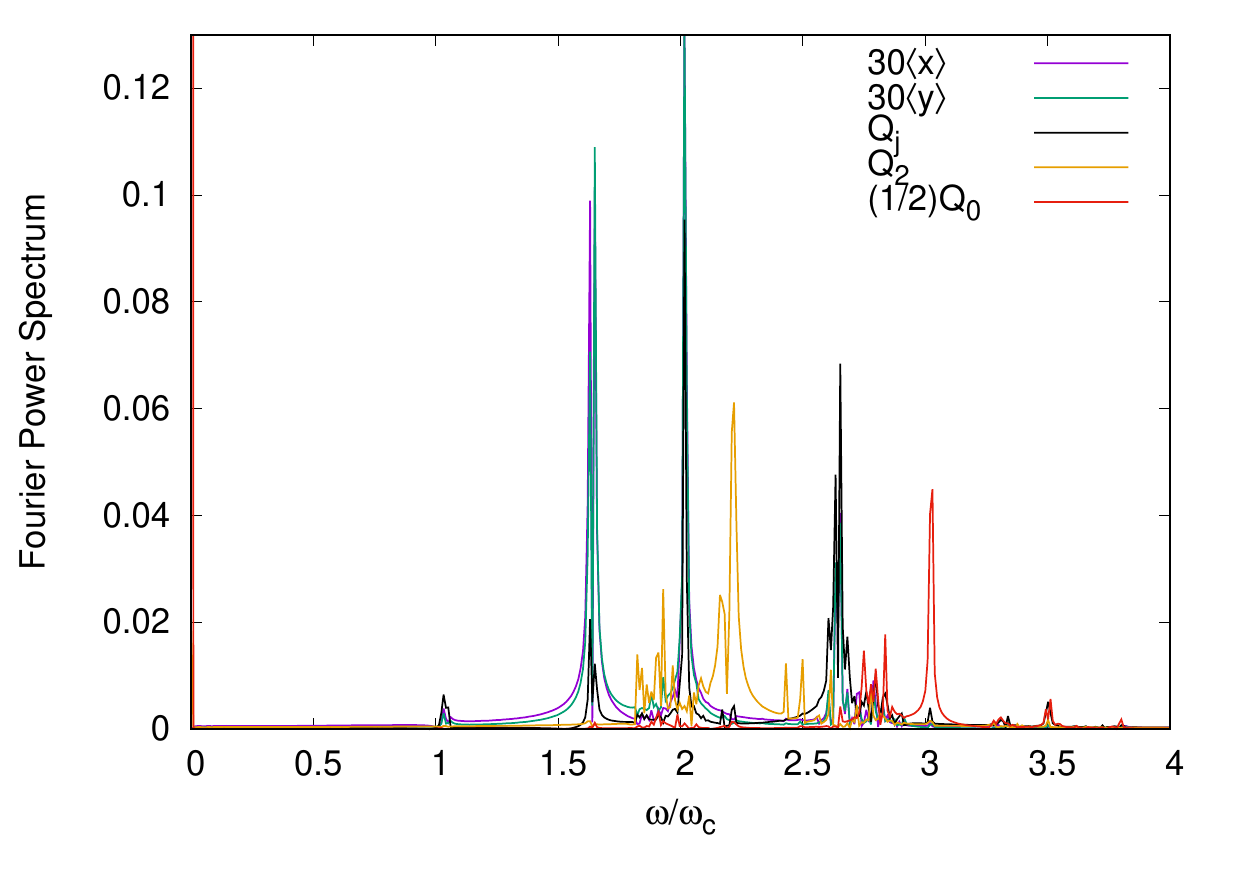}}
	\caption{The Fourier power spectra for a rectangular array of quantum dots for $pq=4,\;
	         c_r=+1$ (left, top),
	         $pq=3,\; c_r=+1$ (right, top), $pq=4,\; c_r=-1$ (left, bottom), and $pq=3,\; c_r=-1$
	         (right, bottom) for $N_e=1$, $k_xL=k_yL=0.785$, $V_\mathrm{t}=0.001$ meV, $V_0=-16$ meV,
	         $L=100$ nm, $T=1.0$ K, $\hbar\Gamma = 1.5$ meV, and $\hbar\Omega = 4.5$ meV.}
	\label{FigQD-pq4-3}
\end{figure*}

This is done in Fig.\ \ref{FigQD-pq4-3} for an excitation pulse with circular polarization
(\ref{phi-ext}) as it is particular well suited to excite collective current modes
in an external magnetic field.
In Fig.\ \ref{FigQD-pq4-3} the Fourier power spectra for the expectation values of the monopole operator
$Q_0$, the dipole operators $\langle x\rangle$ $\langle y\rangle$, the quadrupole operator $Q_2$ and
the current operator $Q_j$ are displayed. These operators are all introduced in the Model Section, \ref{Model}, just before and with Eq.\ (\ref{Qj}).
The results are presented for both directions of the circular polarization, $c_r=\pm 1$.
For $pq=4$ in the left panels of Fig.\ \ref{FigQD-pq4-3} we note the $Q_1$
($\langle x\rangle$ and $\langle x\rangle$) Kohn peaks at
$\omega/\omega_c\approx 1.18$ and 2.18 and in the same location peaks representing current excitations,
$Q_j$. Exactly in between the Kohn peaks is a peak for $Q_j$ at $\omega/\omega_c\approx 1.68$,
below which is a strong quadrupole peak at $\omega/\omega_c\approx 1.5$. Clearly, the strength of the
$Q_j$ peaks depends on the direction of the circular polarization $c_r$. The main $Q_j$ peak at
$\omega/\omega_c\approx 1.68$ is essentially a manifestation of the cyclotron resonance
in a quantum dot made possible by the breaking of the Kohn theorem by the confinement potential
and the external impulse delivered by the excitation pulse. The location of the lower quadrupole
peak (the one seen here) is in accordance with the results shown in Fig.\ 4 in
Ref.\ \cite{Gudmundsson91:12098}.

Wilson et al.\ published a simple classical model of the cyclotron resonances of
a parabolically confined electron in a constant magnetic field, in an article exploring
possible electron phases in an Si inversion layer in a strong magnetic field
\cite{PhysRevB.24.5887}. They identify two cyclotron modes, the low energy anticyclotron
mode, $\omega_1$, made possible by the parabolic confinement potential, and the normal cyclotron resonance, $\omega_2$. With our parameters here, estimating $\hbar\omega_0\approx 6.0$ meV,
this classical model for the flux $pq=4$ gives $\omega_1/\omega_c\approx 1.66$ and $\omega_2/\omega_c\approx 2.66$. The lower mode is very close to the main cyclotron mode seen
in the left panels of Fig.\ \ref{FigQD-pq4-3}, and for $pq=3$ the lower peak is at $\omega_1/\omega_c\approx 2.34$  compared to $\approx 2$ in the right panels of
Fig.\ \ref{FigQD-pq4-3}. The larger deviation for the lower flux, $pq=3$, is expected
as then the effective confinement is farther away from the parabolic case.
The higher mode, the usual cyclotron mode, of the classical model does probably have a
low strength due to the rather high confinement that makes the classical model inadequate, and
we notice instead that both the Kohn peaks have a contribution from the $Q_j$ excitation.
The confinement potential and the magnetic field are mixing the purely longitudinal and
transverse collective modes.

The right panels of Fig.\ \ref{FigQD-pq4-3} show the Fourier power spectra for $pq=3$. Still we can
identify the Kohn dipole peaks and the quadrupole peak, that is now located just above the cyclotron
resonance peak at $\omega /\omega_c\approx 2$.

Important is to notice that generally the cyclotron, the quadrupole, and the monopole modes are
much stronger activated by the circularly polarized excitation pulse than the dipole modes.
This reflects the strong role of the external magnetic field.

In a usual setup of FIR-absorption experiments the wavevector or the impulse is
very small as the wavelength of the radiation is much larger than the lattice
length $L=100$ nm, but in Raman scattering measurements and in modeling thereof the system
is excited with a larger wavevector
\cite{DAHL1994441,Steinebach99:10240,Steinebach00:15600,PhysRevB.59.14892},
as the radiation is inelastically scattered of the electron system.

\begin{figure*}[htb]
  	{\includegraphics[width=0.47\textwidth]{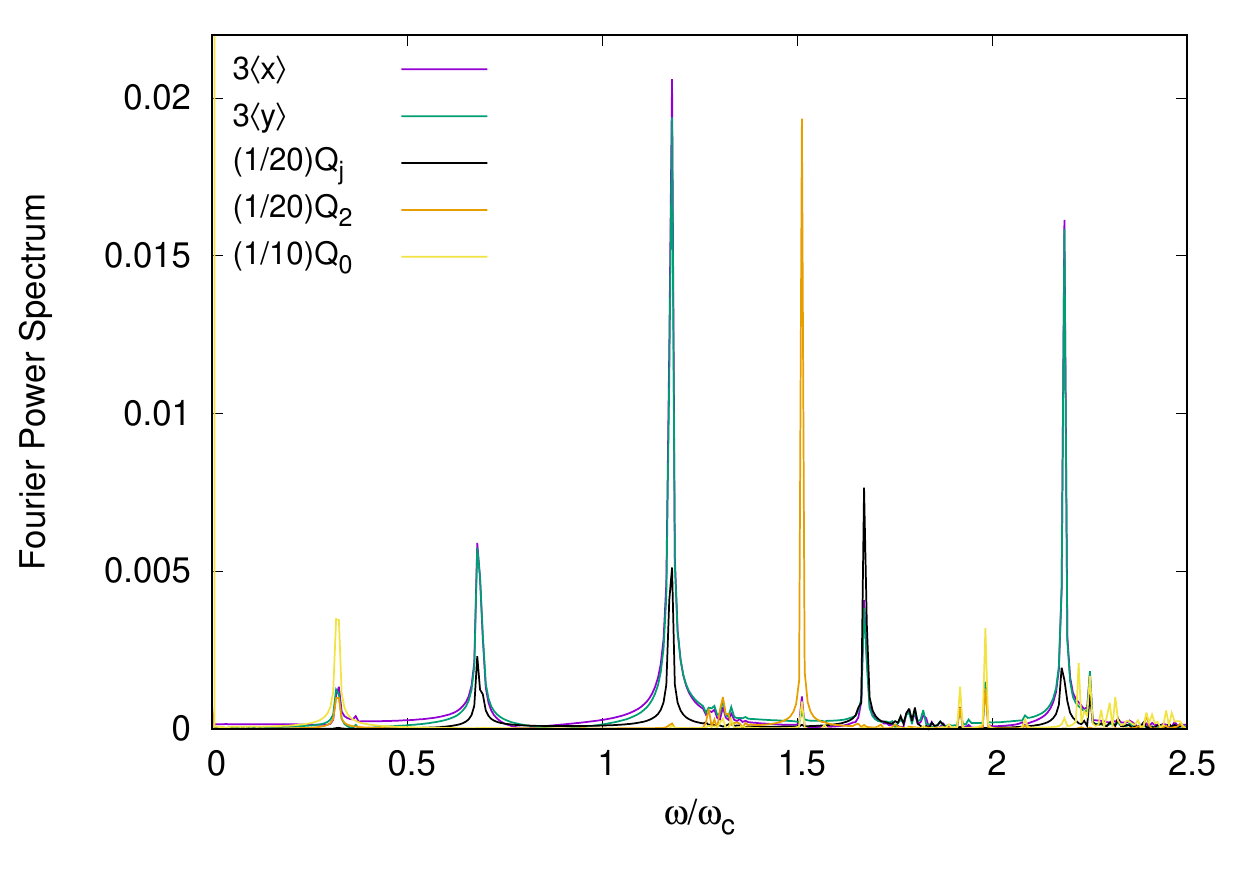}}
	{\includegraphics[width=0.47\textwidth]{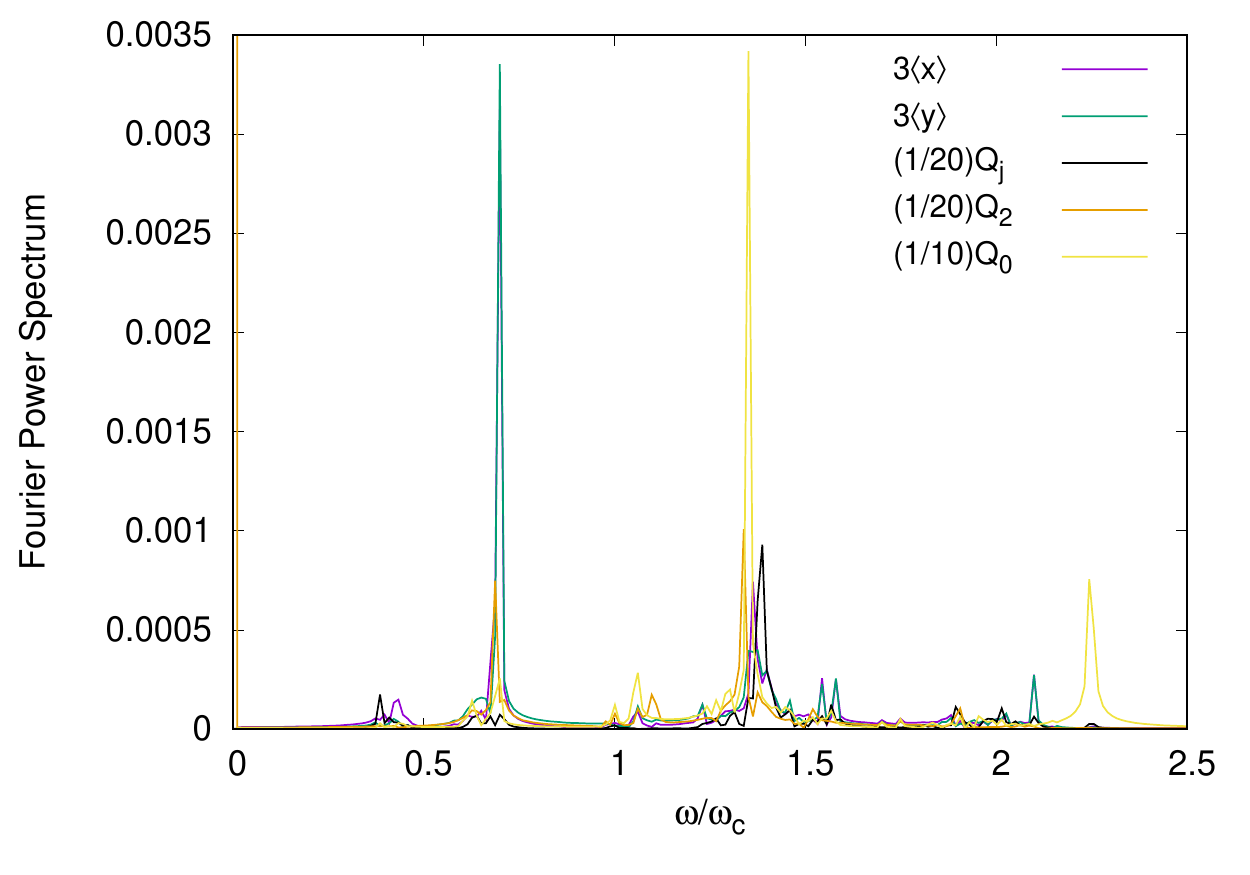}}\\
	{\includegraphics[width=0.47\textwidth]{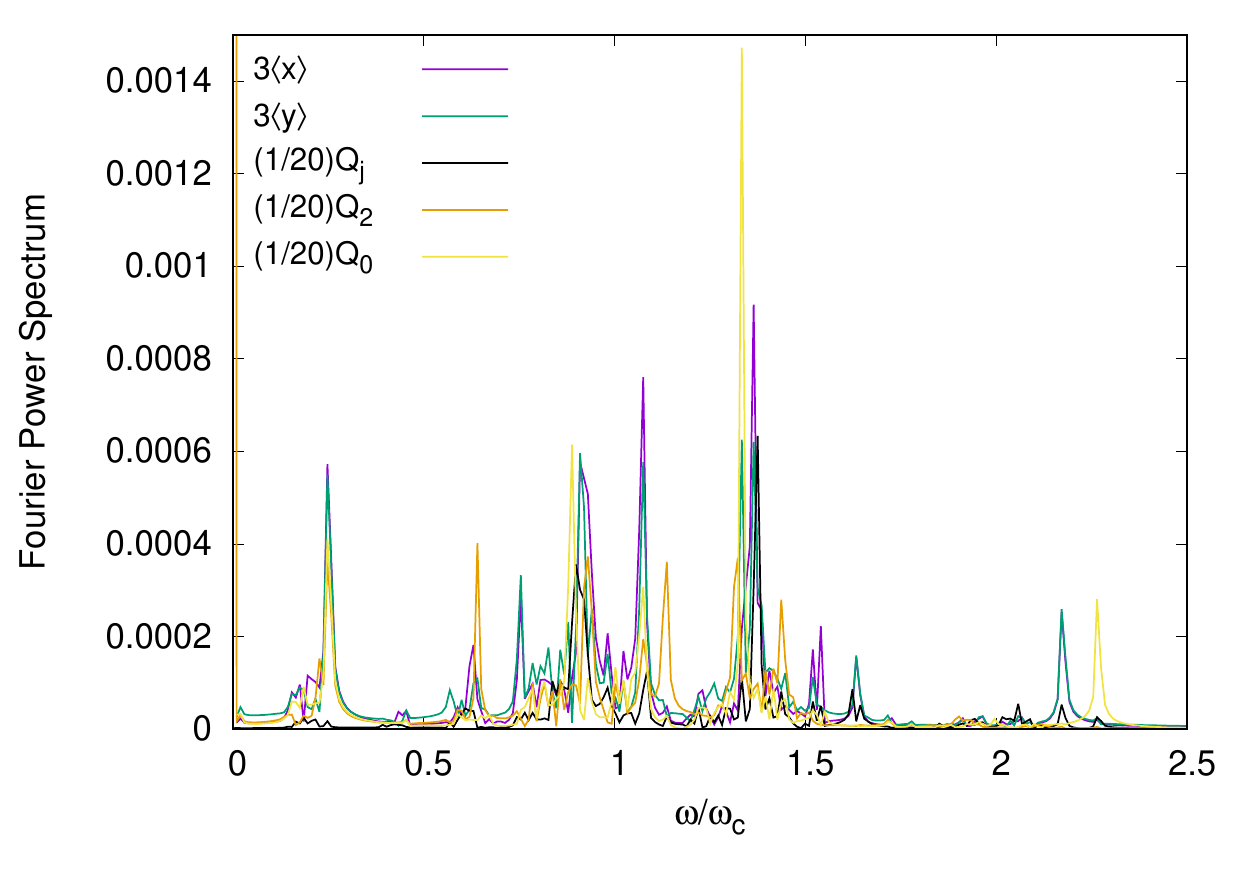}}
	{\includegraphics[width=0.48\textwidth]{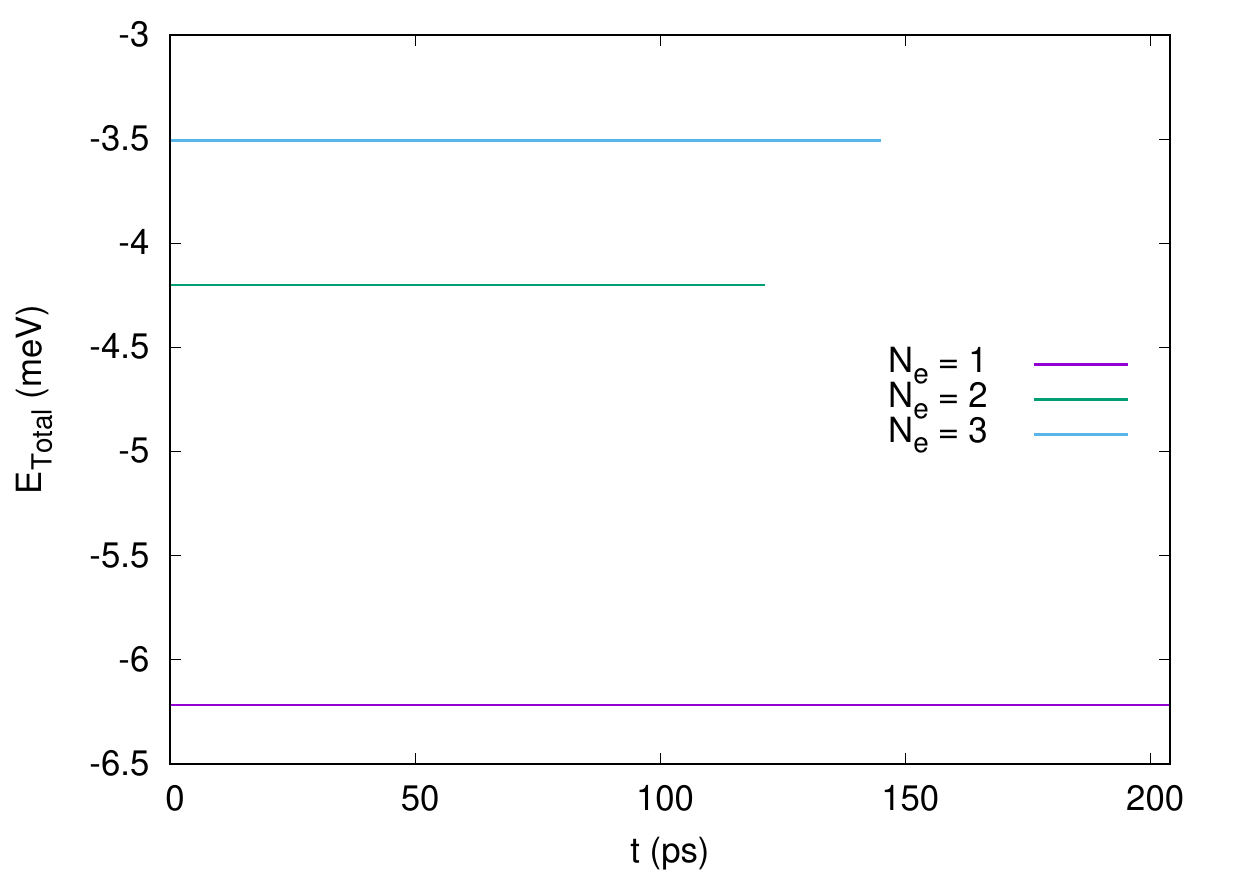}}
	\caption{The Fourier power spectra for a rectangular array of quantum dots for $pq=4$ and
	         linearly polarized excitation field for $N_e=1$ and $V_\mathrm{t}=0.01$ meV
	         with $k_xL = 0.785$ and $k_yL = 0$ (left top),
	         $N_e=2$ and $V_\mathrm{t}=0.001$ meV with $k_xL = 0.524$ and $k_yL = 0$ (right top),
	         $N_e=3$ and $V_\mathrm{t}=0.001$ meV with $k_xL = 0.524$ and $k_yL = 0$ (left bottom),
	         and the average energy of the system
	         (right bottom). $V_0=-16$ meV, $L=100$ nm, $T=1.0$ K, $\hbar\Gamma = 1.5$ meV,
	         and $\hbar\Omega = 4.5$ meV.}
	\label{Fig-pq4-lin-varNe}
\end{figure*}

A single electron in an isolated quantum dot would not be well described by the
Hartree approximation, but here is important to have in mind that one electron in a
quantum dot feels the influence of the electrons in the neighboring unit cells.
With our selected parameters each electron feels the influence of 25 electrons.
Its self-interaction is reduced by the positive charge background representing the
ions in the crystal. Before going to lower magnetic fields we show in
Fig.\ \ref{Fig-pq4-lin-varNe} how the excitation spectrum changes as the number
of electrons is increased in each dot from 1 to 3 and in the bottom right panel
how the mean total energy changes. The mean total energy can be compared to the
confining potential in Fig.\ \ref{VxyApprox} in order to see how the symmetry
of the underlying square lattice affects the potential as the number of electrons
increases in the dots. Not shown here, but for 1 or 2 electrons in a dot the overlap of the
probability density is very small between neighboring dots, but for the case of 3 electrons
the density between the dots reaches 10\% of its peaks value.

\begin{figure*}[htb]
	{\includegraphics[width=0.47\textwidth]{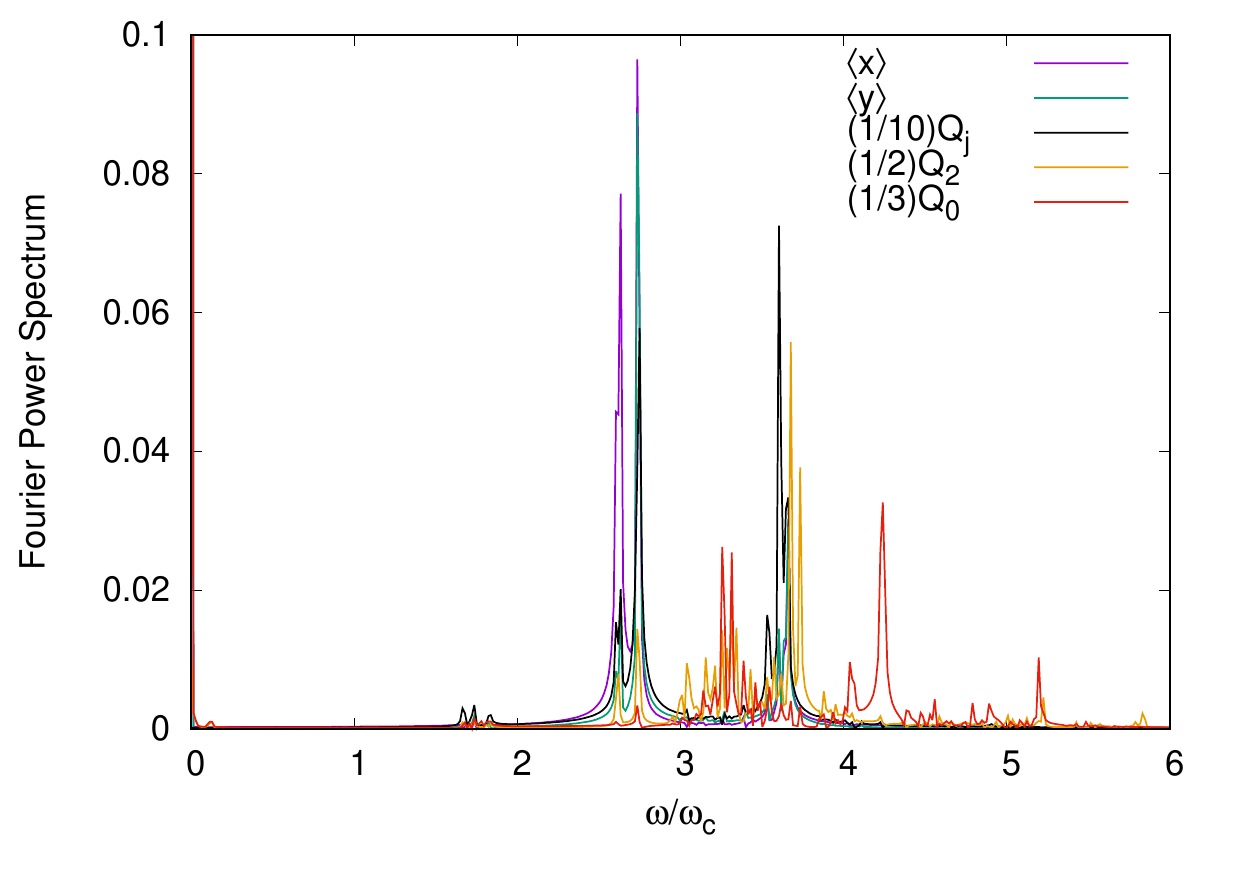}}
	{\includegraphics[width=0.47\textwidth]{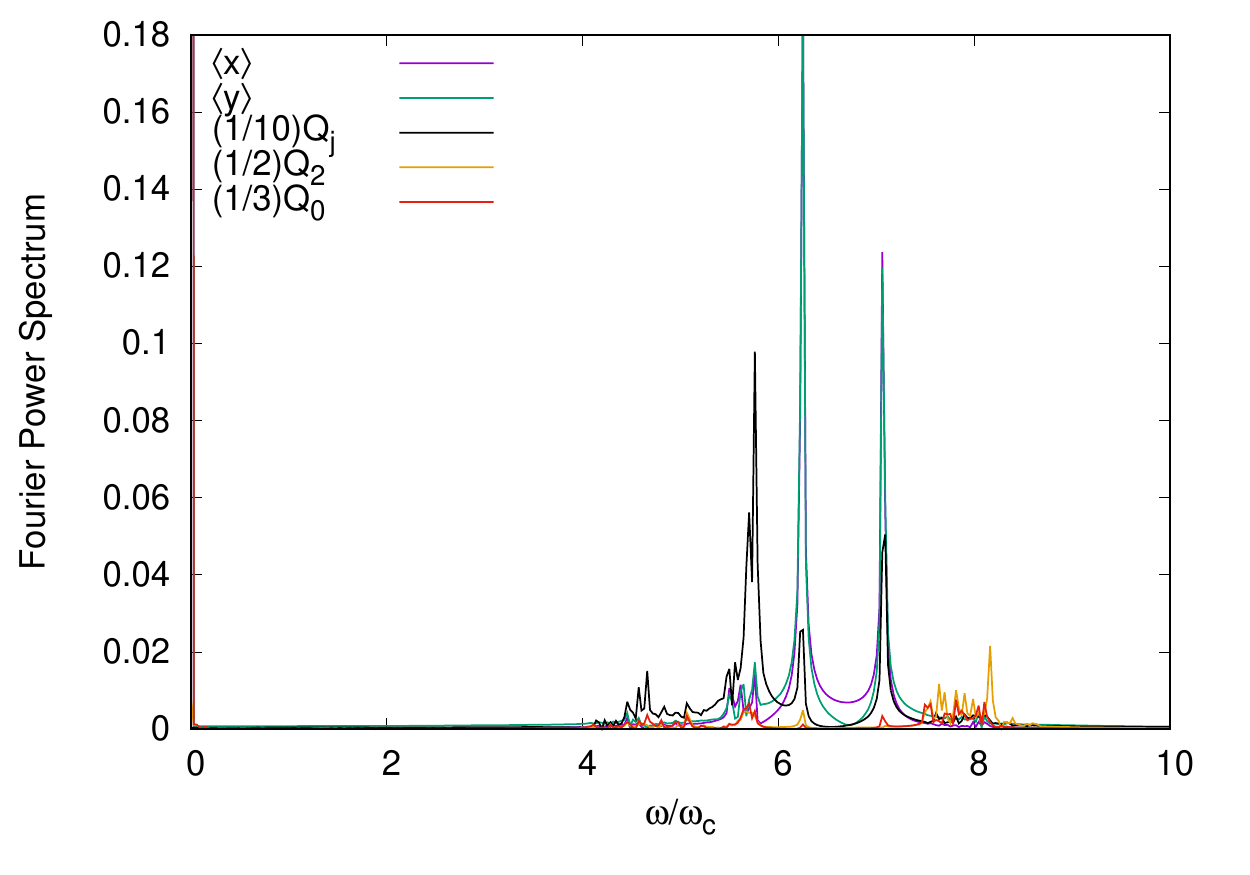}}\\
	{\includegraphics[width=0.47\textwidth]{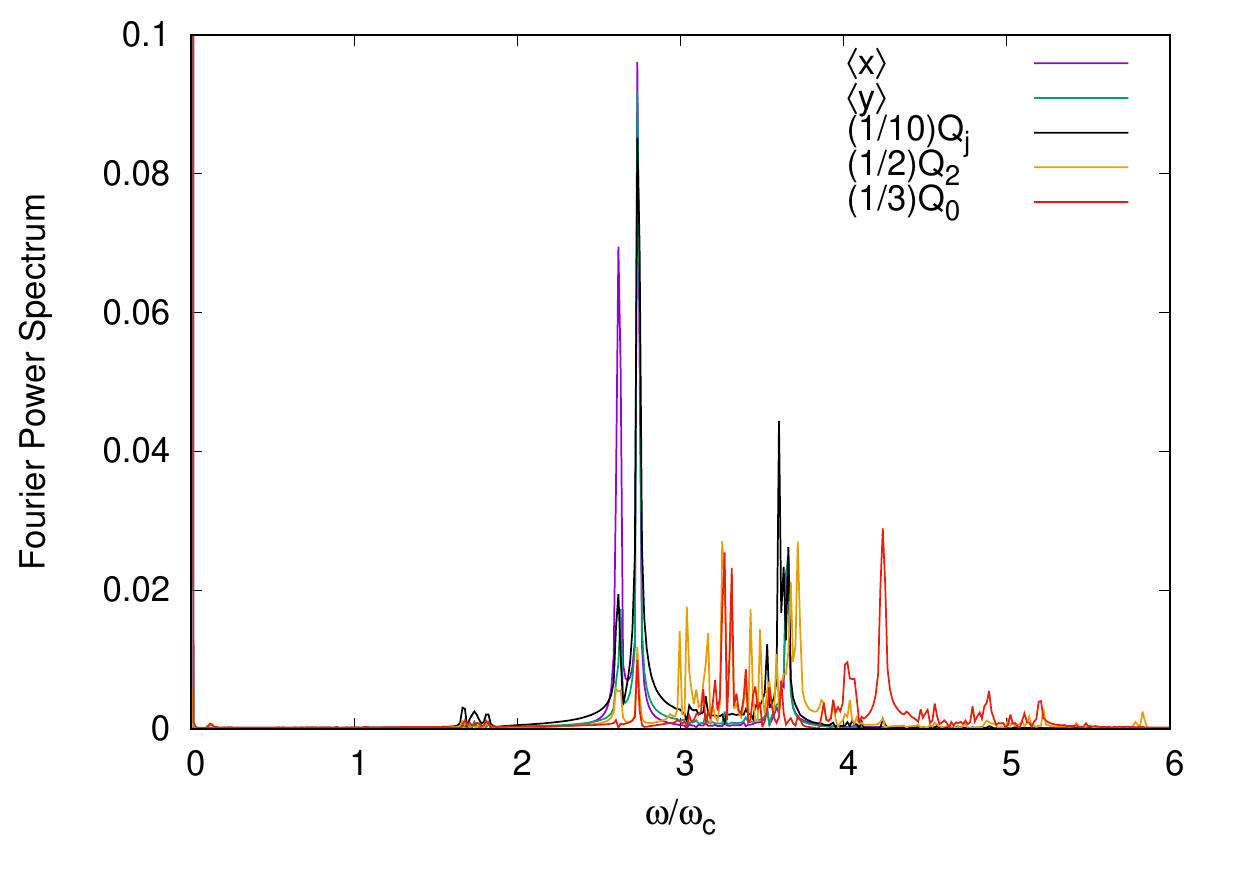}}
	{\includegraphics[width=0.47\textwidth]{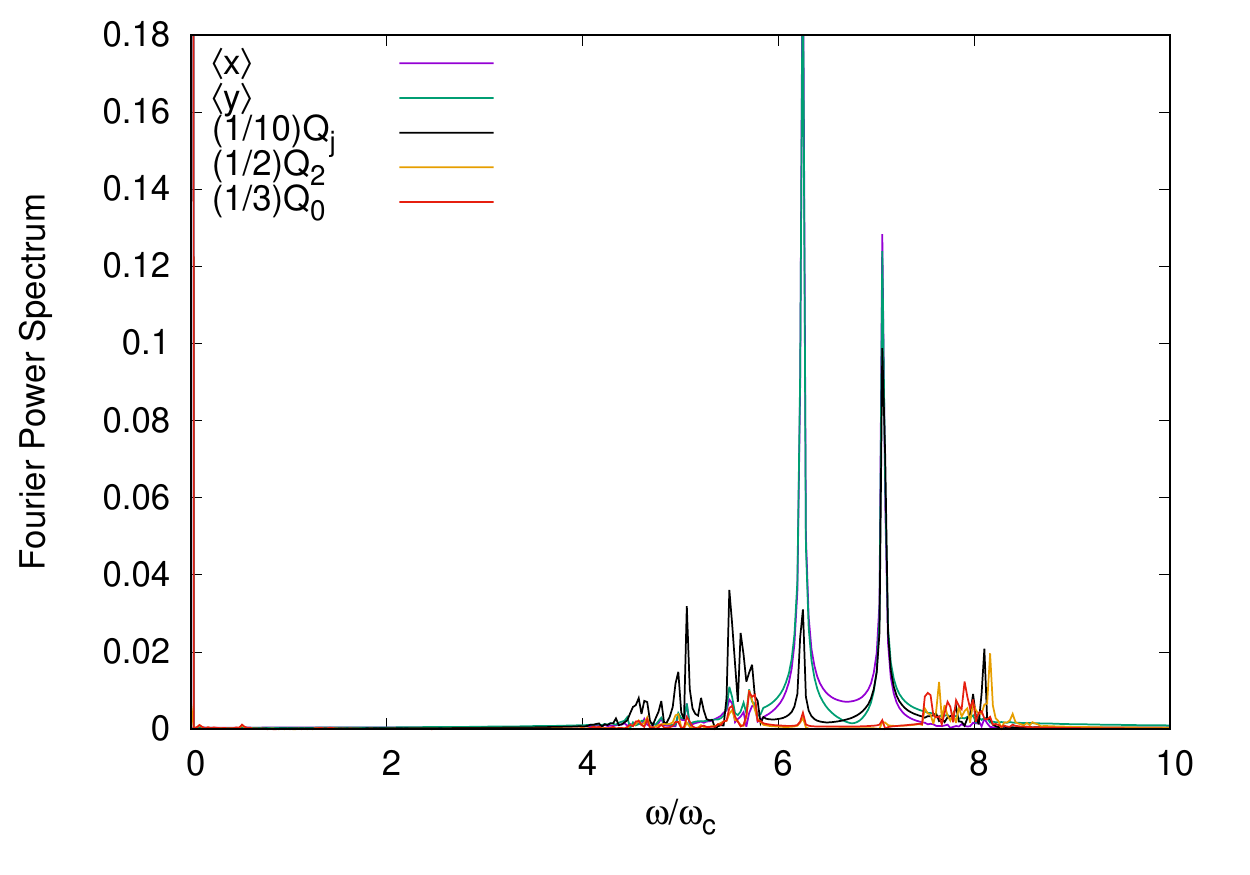}}
	\caption{The Fourier power spectra for a rectangular array of quantum dots for $pq=2,\; c_r=+1$ (left, top),
	         $pq=1,\; c_r=+1$ (right, top), $pq=2,\; c_r=-1$ (left, bottom), and $pq=1,\; c_r=-1$
	         (right, bottom) for $N_e=1$, $k_xL=k_yL=0.785$ for $pq=2$, but $k_xL=k_yL=0.393$ for $pq=1$,
	         $V_\mathrm{t}=0.001$ meV, $V_0=-16$ meV, $L=100$ nm, $T=1.0$ K,
             $\hbar\Gamma = 1.5$ meV, and $\hbar\Omega = 4.5$ meV.}
	\label{FigQD-pq2-1}
\end{figure*}

We continue with the Fourier power spectra for quantum dots with $N_e=1$
shown in Fig.\ \ref{FigQD-pq4-3}
for a circularly polarized excitation pulse and observe in Fig.\ \ref{FigQD-pq2-1} the results
for $pq=2$ (left panels) and $pq=1$ (right panels). For the lower magnetic flux ($pq=1$) we
see two clear Kohn peaks, where the lower one is clearly split due to the square symmetry
of the lattice, and the cyclotron resonance is below the Kohn peaks.
For the higher magnetic flux ($pq=2$) the lower Kohn peak is difficult to resolve clearly due to a
fine splitting caused by the lattice and both Kohn peaks interact strongly with cyclotron resonance peaks.
Interestingly, the monopole peaks are prominent.

\subsection{The limit to a flat system}
\begin{figure}[htb]
	{\includegraphics[width=0.47\textwidth]{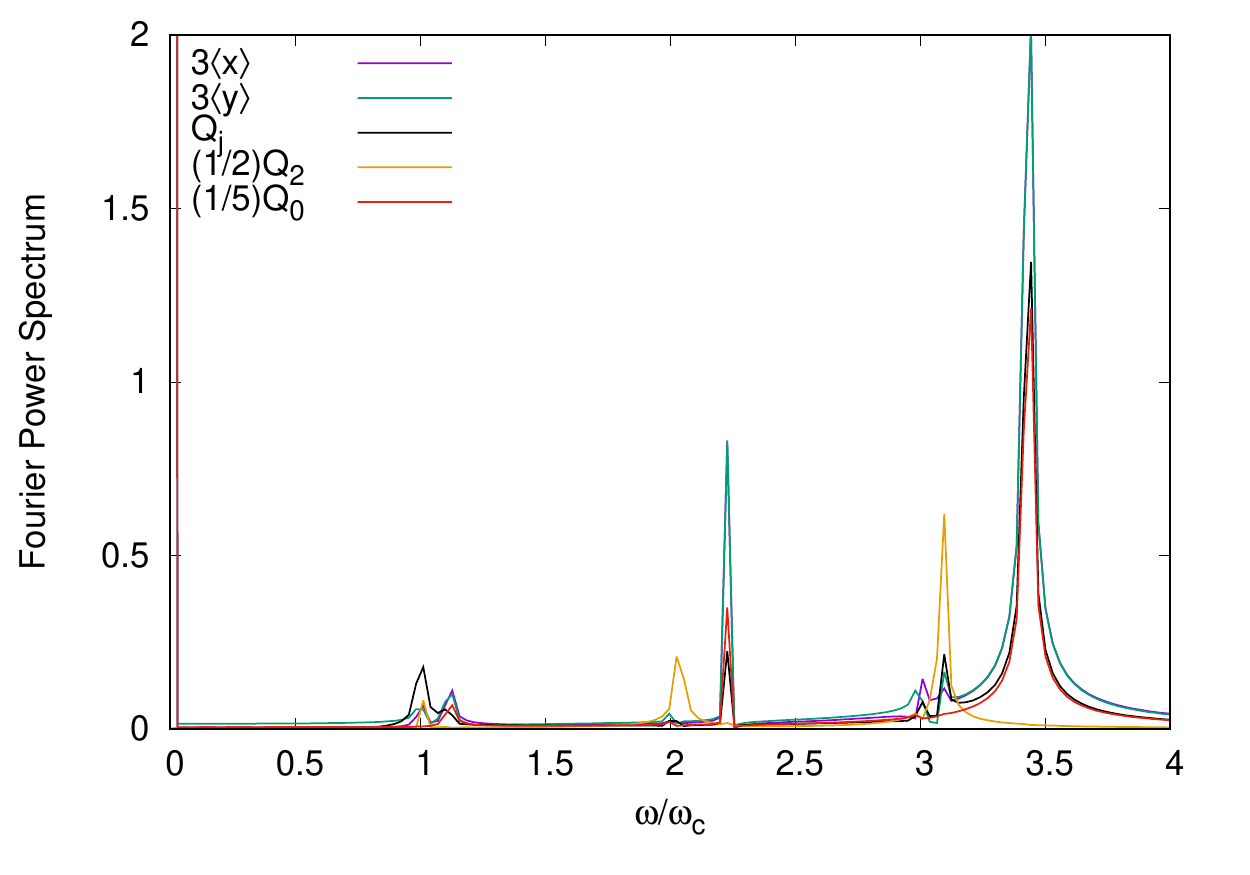}}
	{\includegraphics[width=0.47\textwidth]{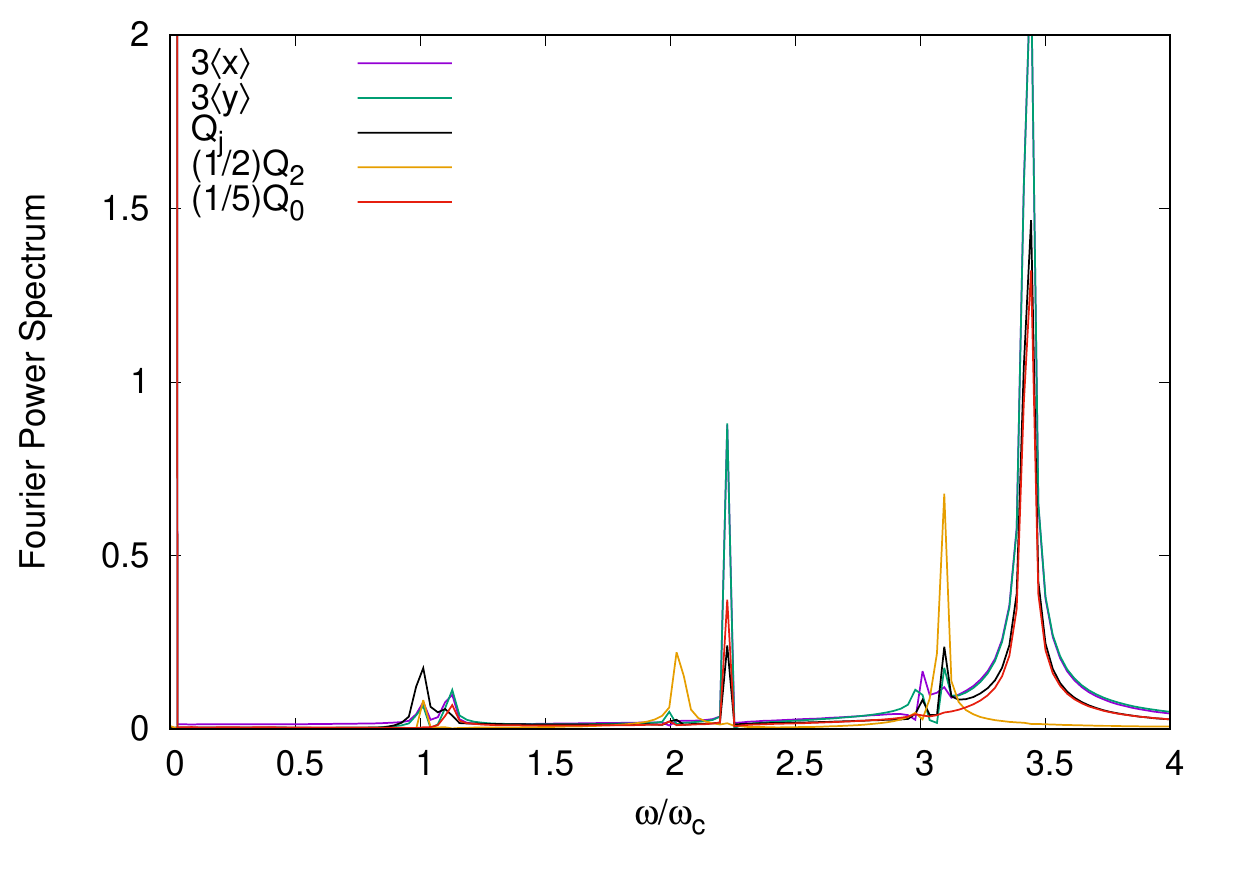}}
	\caption{The Fourier power spectra for electrons in a rectangular array of flat unit cells for
	         $pq=1,\; c_r=+1$ (top), $pq=1,\; c_r=-1$ (bottom) for $N_e=1$, $k_xL=k_yL=10^{-4}$,
	         $V_\mathrm{t}=0.01$ meV, $V_0=0$ meV, $L=100$ nm, $T=0.1$ K,
             $\hbar\Gamma = 1.5$ meV, and $\hbar\Omega = 4.5$ meV.}
	\label{FigFl-pq1}
\end{figure}

The presence of the cyclotron resonance peaks in the excitation spectra for the arrays of quantum
dots awakes the question: What happens when the strength of the modulation defining the dot array,
$V_0\rightarrow 0$? Important is to have in mind that even though the modulation is set to vanish
the dynamic of the system is not set totally free as the restrictions of periodicity are still imposed
on the system in the model. In order to start with results of some familiarity we display the results
for the lowest magnetic flux $pq=1$ for $V_0=0$ in Fig.\ \ref{FigFl-pq1}.
There are small cyclotron resonance peaks at $\omega /\omega_c = 1$, 2, and 3 as expected and for
a slightly higher energy split plasmon peaks follow. This situation reminds us of the absorption
spectra shown in Fig.\ 1 in Ref.\ \cite{Gudmundsson96:5223R} for linear response. There we see dipole
plasmon peaks with increasing oscillator strengths in higher Landau levels as the impulse $kL$ increases
with Bernstein-type splitting \cite{Bernstein58:10,Gudmundsson95:17744}. Here, we can not identify
Bernstein modes as $k_xL=10^{-4}$ and $k_yL=10^{-4}$ is very low, but due to a higher resolution
we can identify higher order plasmon modes. In addition, here
we see the cyclotron resonances and further plasmon details as the quadrupole and monopole modes.
Even though the excitation pulse here carries a small impulse the periodicity of the systems
represents a larger possible impulse, and as the Landau level separation for $pq=1$ is small
the excitation pulse may couple better to higher cyclotron resonances.

\begin{figure*}[htb]
	{\includegraphics[width=0.47\textwidth]{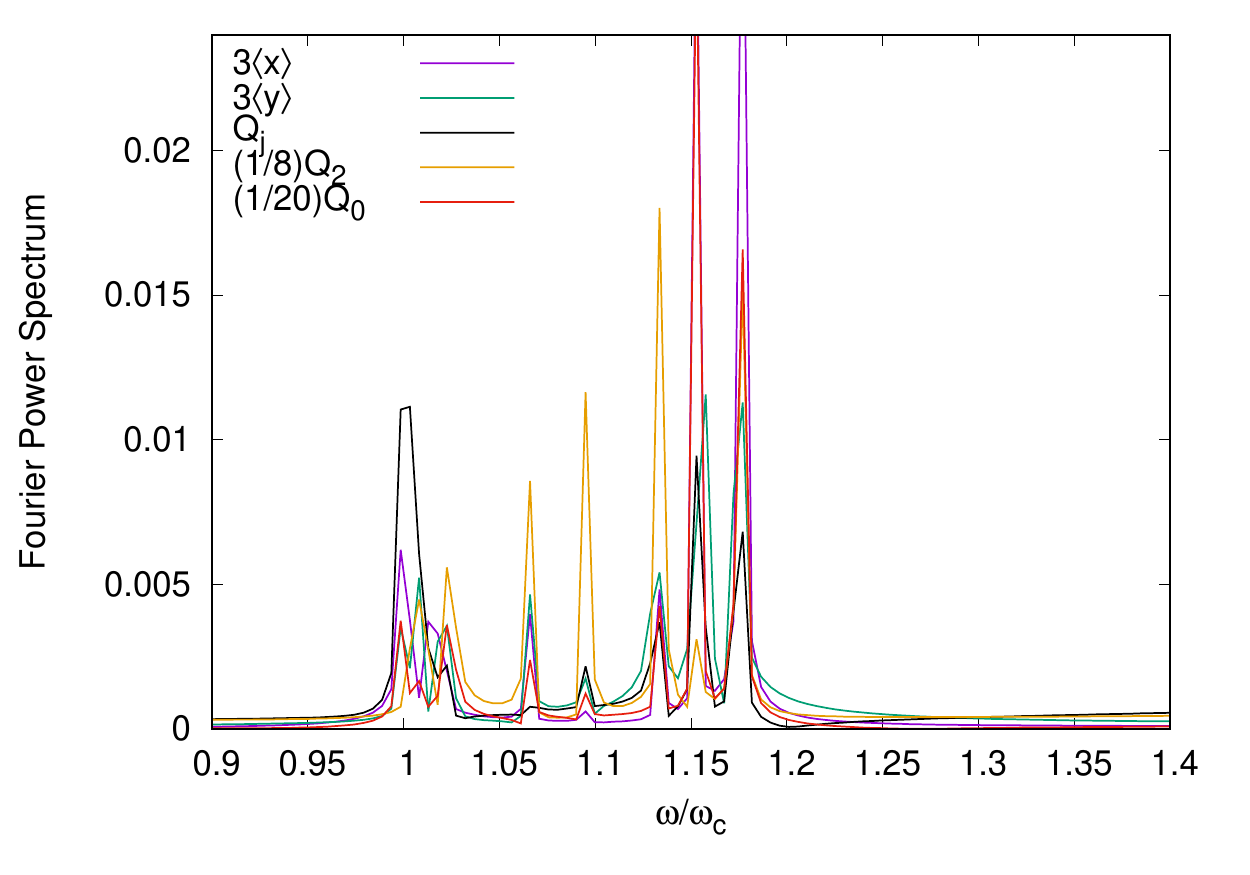}}
	{\includegraphics[width=0.47\textwidth]{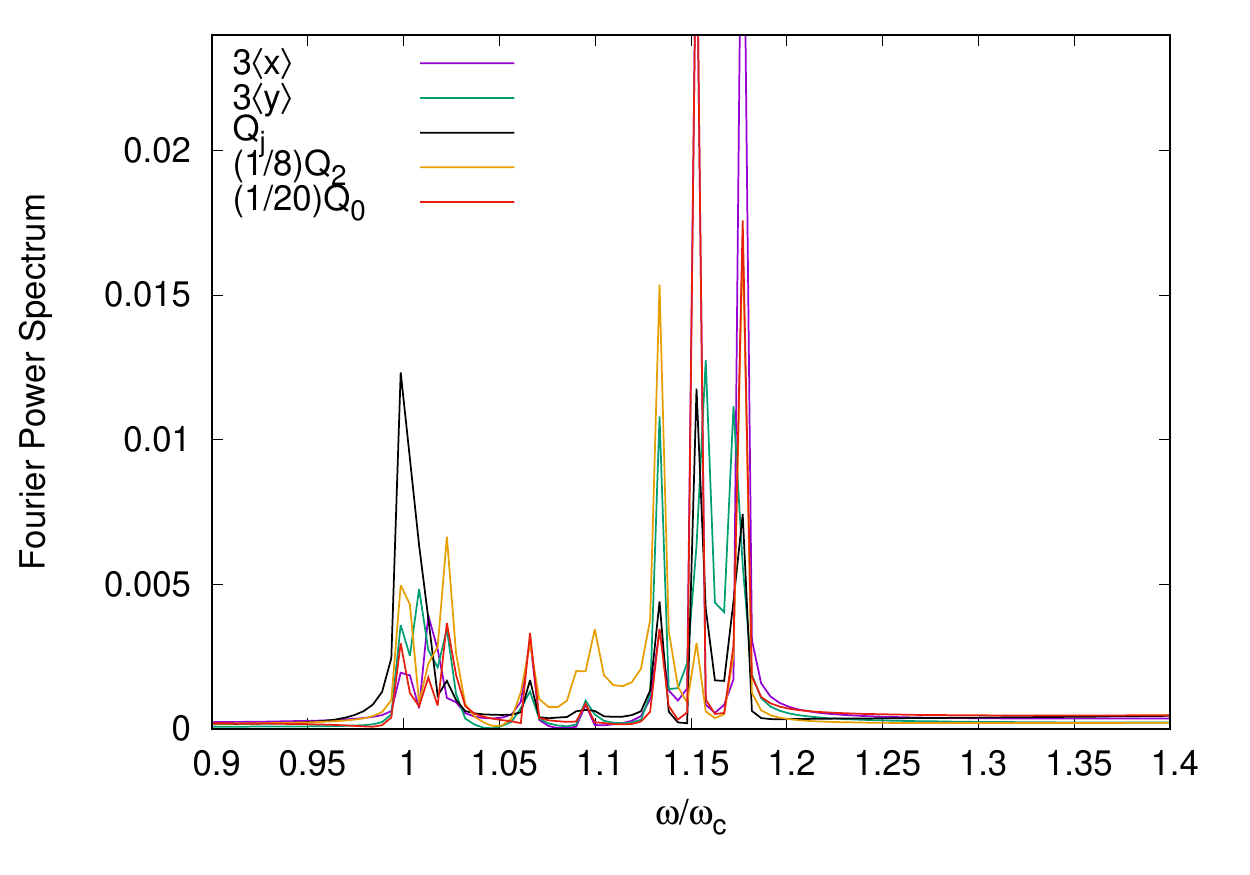}}\\
	{\includegraphics[width=0.47\textwidth]{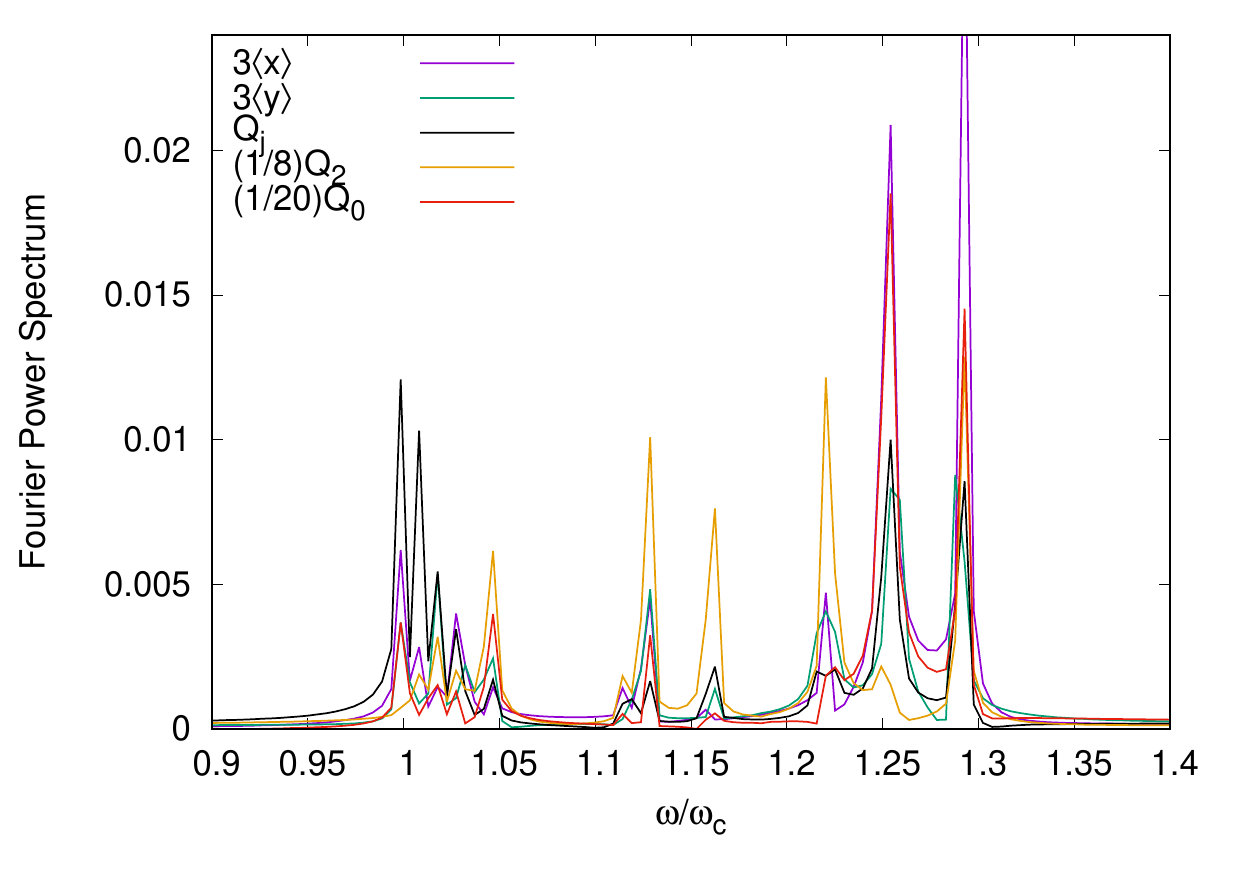}}
	{\includegraphics[width=0.47\textwidth]{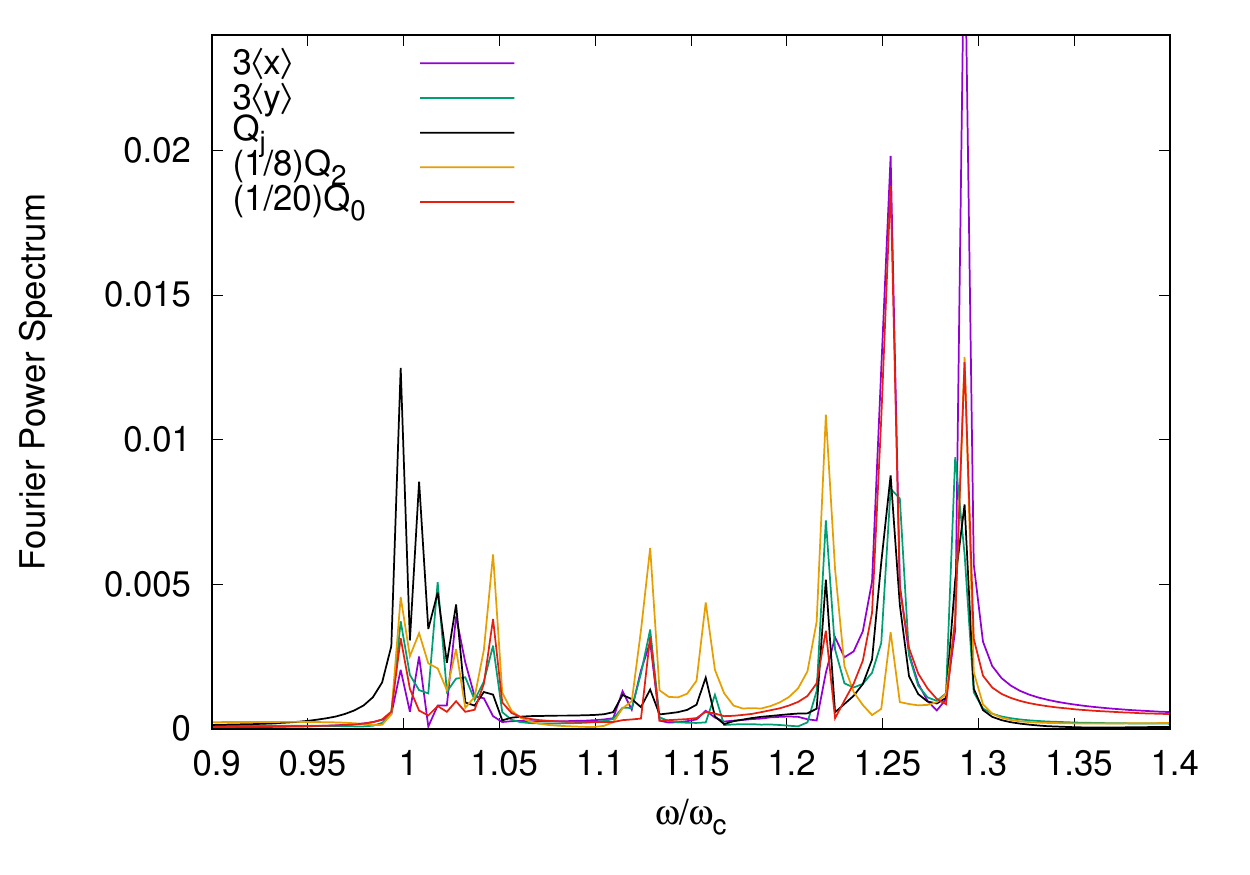}}\\
	{\includegraphics[width=0.47\textwidth]{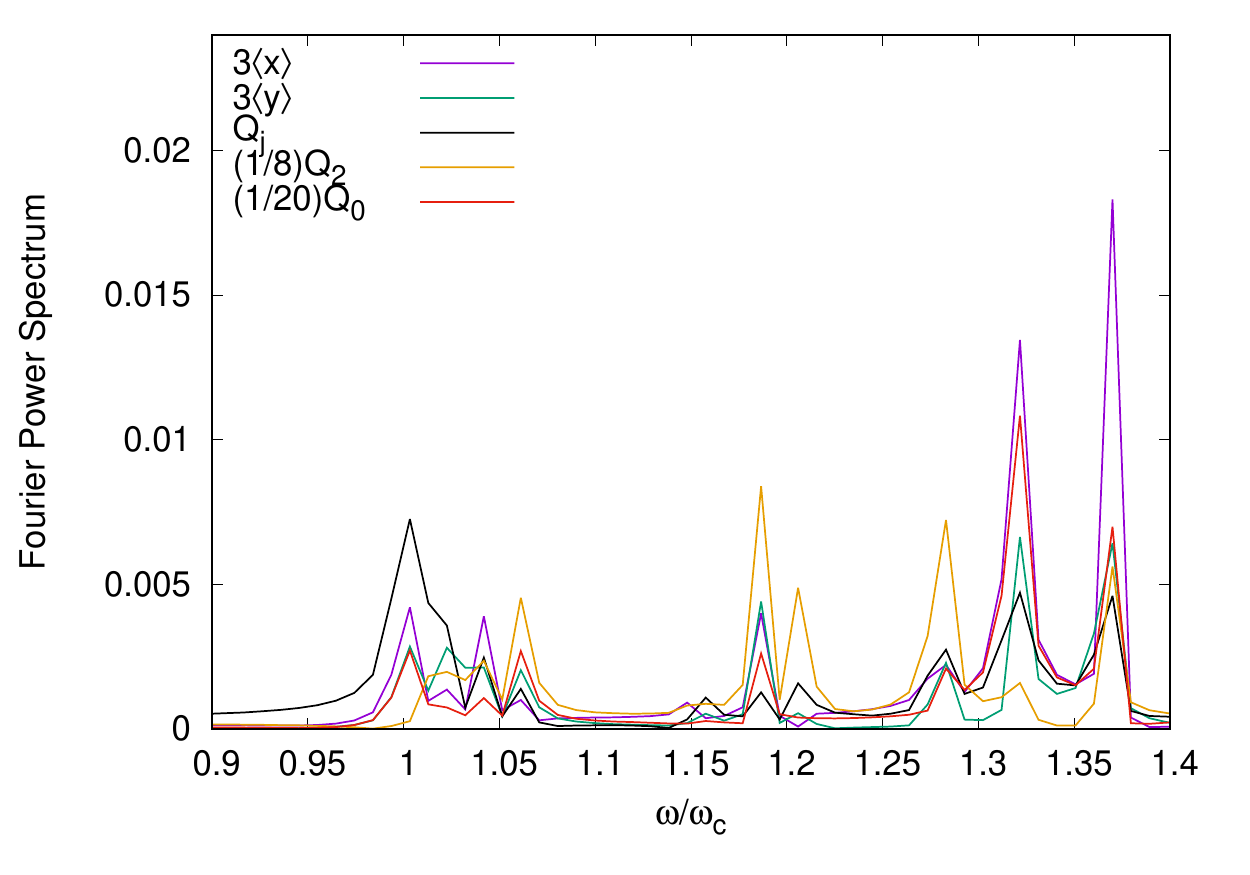}}
	{\includegraphics[width=0.47\textwidth]{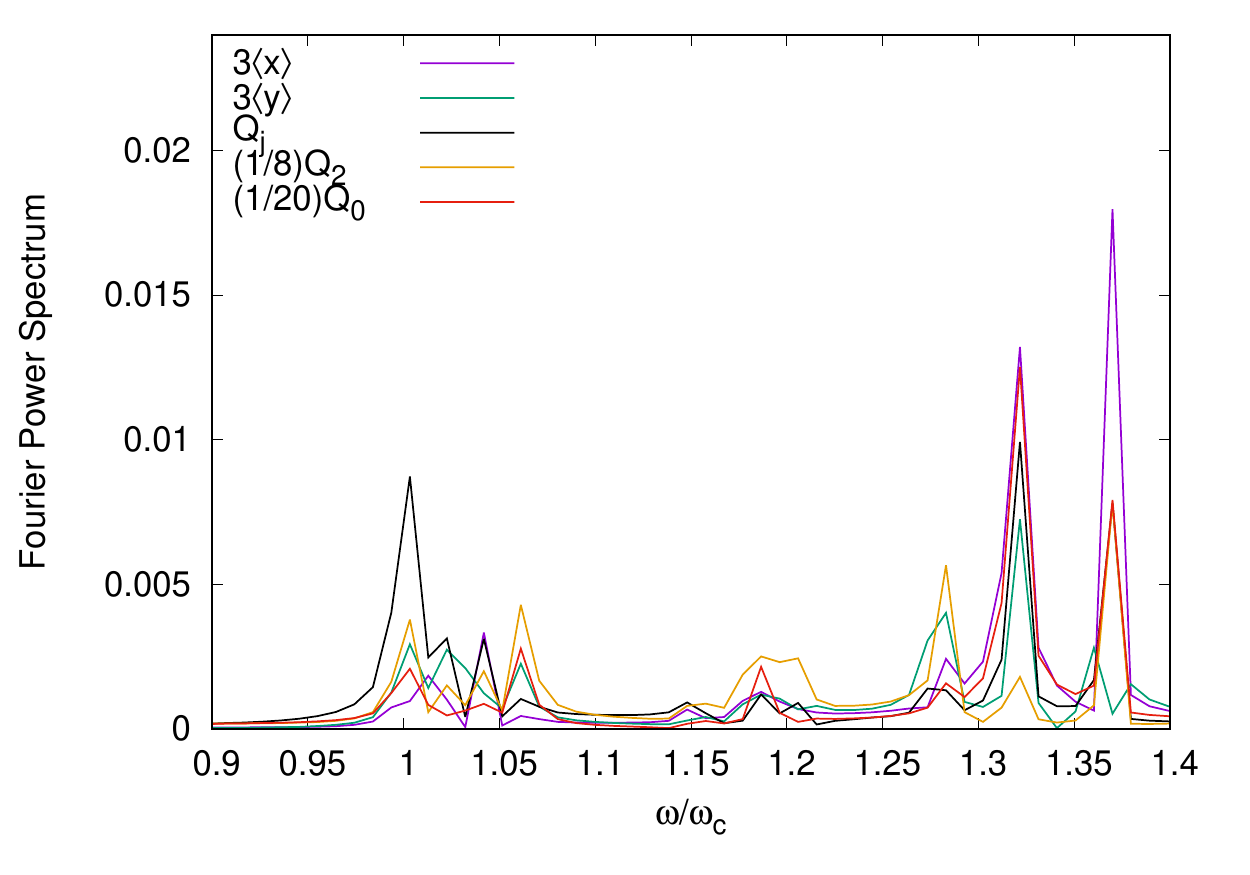}}
	\caption{The Fourier power spectra for electrons in a rectangular array of flat unit cells for
	         $c_r=+1,\; N_e =1$ (left, top), $c_r=-1\; N_e=1$ (right, top),
	         $c_r=+1,\; N_e =2$ (left, center), $c_r=-1\; N_e=2$ (right, center),
	         $c_r=+1,\; N_e =3$ (left, bottom), $c_r=-1\; N_e=3$ (right, bottom)
	         for $pq=3$, $k_xL=k_yL=0.785$, $V_\mathrm{t}=0.001$ meV, $V_0=0$ meV, $L=100$ nm, $T=1.0$ K,
             $\hbar\Gamma = 1.5$ meV, and $\hbar\Omega = 4.5$ meV.}
	\label{FigFl-pq3-NeVar}
\end{figure*}

In Fig.\ \ref{FigFl-pq3-NeVar} the Fourier power spectra for the higher flux $pq=3$ are shown for
the range $0.9\leq\omega /\omega_c\leq 1.4$ for a periodic system with vanishing modulation
$V_0=0$ for different number of electrons $N_e=1$ (top), 2 (center), 3 (bottom) and both circular
polarizations $c_r=+1$ (left) and -1 (right). For this vanishing modulation at $pq=3$ each Landau
level is 3-fold degenerate.
The spectra in Fig.\ \ref{FigFl-pq3-NeVar} all show a clear cyclotron fundamental resonance peak with
a maximum at $\omega /\omega_c =1$, and all show a set of three-fold plasmon peaks at a higher energy
before they become flat. As expected, the plasmon peaks move to higher energy as the number of electrons
in a unit cell increases. The structure of spectra is more complex, just above the first cyclotron
resonance at $\omega /\omega_c =1$ there are three-fold peaks, and in between the main set of peaks
mentioned here, there are 2 or 3 peaks still with high quadrupole contribution.

Not shown here, but for the higher flux $pq=4$ we see sets of 3 - 4 peaks appearing in the
excitation spectra. Possibly, longer time series for the Fourier analysis would result in clear
sets of 4 peaks.

\begin{figure}[htb]
	{\includegraphics[width=0.48\textwidth]{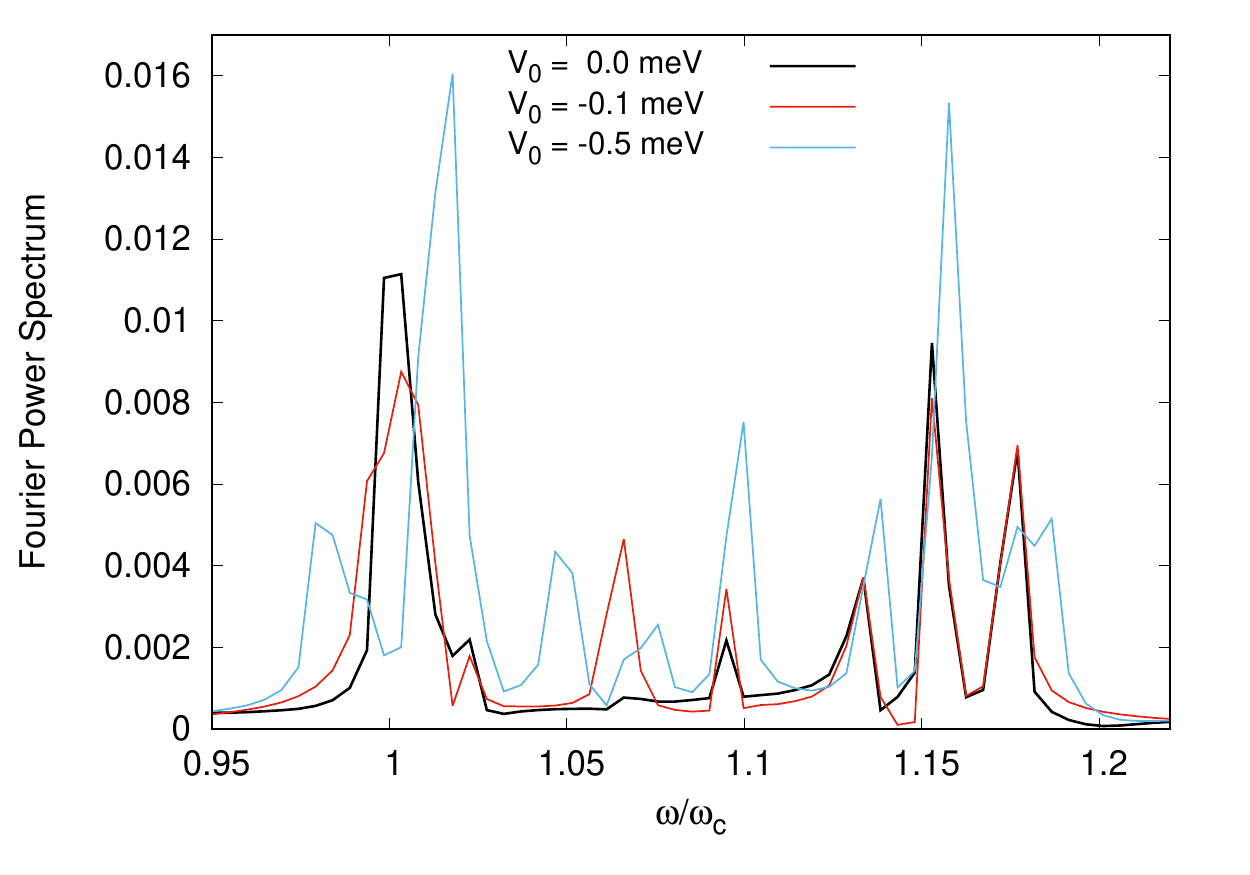}}
	\caption{The Fourier power spectra for current excitations $Q_j$ for electrons in a rectangular array
	         of flat or slightly modulated unit cells with a dot potential for varying degree
             of modulation $V_0$. $N_e=1$, $pq=3$, $c_r=+1$, $k_xL=k_yL=0.785$, $V_\mathrm{t}=0.001$ meV,
             $L=100$ nm, $T=1.0$ K, $\hbar\Gamma = 1.5$ meV, and $\hbar\Omega = 4.5$ meV.}
	\label{Fig-varV0-pq3}
\end{figure}

In Fig.\ \ref{Fig-varV0-pq3} the excitation spectrum for the current excitations $Q_j$ corresponding to
the upper left panel of Fig.\ \ref{FigFl-pq3-NeVar} is repeated for vanishing modulation $V_0=0$
together with the results for $V_0=-0.5$ meV and -0.1 meV in order to show the stability of the
spectra as the modulation approaches 0.
Especially, the 3-fold plasmon peak at the higher energy end of the spectrum is insensitive to the
change in the modulation.

\section{Conclusions}
\label{Conclusions}
We have demonstrated that a ``real-time'' excitation of a modulated 2DEG in a constant perpendicular
external magnetic field can be used to explore a wide range of collective modes in the system
including longitudinal and transverse modes. This can not be achieved by the linear response
approach if only density-density correlations are considered, but current-current response
can lead to the emergence of transverse modes. The external magnetic field with its Lorentz
force makes the transverse modes particularly important in the system investigated here.
Furthermore, the real-time approach opens the possibility to explore modes beyond the linear response, i.e.\ nonlinear response, or calculations for pump-and-probe schemes,
though that was not done here.

We are able to attain information about the cyclotron resonances in an array of quantum
dots concurrently with the better known dipole- quadrupole- and monopole plasmon collective
modes. The lack of this information has been pointed out by experimental researchers
for small electronic systems with confinement, that deviates from the parabolic ideal
\cite{Merkt96:1134}.

With the real-time approach we have been able to see how the cyclotron resonances
evolve into their well known form as the confinement potential vanishes in the 2DEG.
In the flat, but periodic, system we see peak structures for the plasmons and the cyclotron
resonances, that reflect the underlying degeneracies of the bandstructure of the
Hofstadter butterfly. We show that this structure of the excitation spectra is rather stable
as the modulation of the 2DEG approaches 0, and can be seen for different numbers of electrons in
the unit cell.

\begin{acknowledgments}
This work was financially supported by the Research Fund of the University
of Iceland, and the Icelandic Infrastructure Fund. The computations were performed on resources
provided by the Icelandic High Performance Computing Center at the University of Iceland.
V.\ Mughnetsyan and V.G. acknowledge support by the Armenian State Committee
of Science (grant No 21SCG-1C012). V.\ Moldoveanu acknowledges financial support from
the Romanian Core Program PN19-03 (contract No.\ 21 N/08.02.2019).

\end{acknowledgments}

%
\appendix

%

\section{Information about the numerical implementation}
\label{AppNumDetails}

In the dynamic Hartree approximation the Hamiltonian for the
system has to be updated in each time-step. At a time $t > 0$
the Hamiltonian is
\begin{equation}
      H(t) = H_0 + V_\mathrm{per} + H^\mathrm{ext}(t) + V_\mathrm{H}[\Delta n(0)]+\delta V_\mathrm{H}[\delta n(t)],
\end{equation}
where Eq.\ (\ref{Ht}) has been elaborated, and
$\delta n(t) = n(t)- n(0)$.
Shortly after, the density can be approximated by
\begin{equation}
      \delta n(t+\delta t) =  n(t+\delta t) - n(0) \approx
      \delta n(t) + \delta n( \delta t)
\end{equation}
for a very short time step $\delta t$ with respect to all time scales in the
system. Thus, the Hamiltonian at the later time will be
\begin{align}
      H(t+\delta t) =& \left\{ H_0 + V_\mathrm{H}[\Delta n(0)] + \delta V_\mathrm{H}[\delta n(t)]\right\}\\ \nonumber
      +&\; V_\mathrm{per} +
      H^\mathrm{ext}(t+\delta t) + \delta V_\mathrm{H}[\delta n(\delta t)],
\end{align}
where the terms in the curly bracket on the right side in the first line
can be considered as the time updated or renormalized Hamiltonian of the
original static system. The effect of the positive homogeneous background charge
of the system, $n_b$, is to cancel out all terms with $\bm{q}=\bm{G}+\bm{k}=0$,
just like is done for the term with $\bm{G}=0$ in the calculation for the static system.

In order to increase the numerical accuracy for the matrix elements of the
dynamical Hartree interaction in Eqs.\ (\ref{V_H}) and (\ref{VHGk}) the Fourier
transform is done after the construction of the variation in the density.
This means that we consider the integrals
\begin{equation}
      \delta n(\bm{G}+\bm{k}) = \int_{\bm{A}} d\bm{x}\:
      e^{i(\bm{G}+\bm{k})\cdot\bm{x}}\: \delta n(\bm{x},\bm{k})
\end{equation}
and
\begin{align}
      \delta n(\bm{x},\bm{k}) = \frac{1}{(2\pi )^2}\sum_{\alpha\beta}\int_{-\pi}^{\pi} d\bm{\theta}\: &
      \psi_{\alpha\bm{\theta}}^*(\bm{x})\psi_{\beta\bm{\theta}+\bm{k}}(\bm{x})\\ \nonumber
      &\Delta\rho_{\beta\bm{\theta}+\bm{k},\alpha\bm{\theta}}.
\end{align}
In addition, this approach simplifies the monitoring of symmetries during the calculations.

Integrations over $\bm{q}=\bm{G}+\bm{k}$ in the reciprocal space is divided into
sums over the inverse lattice vectors $\bm{G}$ and numerical integrations over
$\bm{k}$ with a equispaced grids constructed from repeated applications of
the five point Booles quadrature. The equispaced grid is essential in order to
account for all transitions fulfilling $\bm{q}=\bm{G}+\bm{k}$.
The discreteness of the mesh for $\bm{k}$ and computational costs for
a higher number of points make it difficult to represent the dispersion
of excitation spectra as continuous functions of $\bm{k}$.

%
\frenchspacing

%

%
%
%
\end{document}